\documentclass[12pt]{article}

\usepackage{times}
\usepackage{fixltx2e}
\usepackage{graphicx}
\usepackage[table]{xcolor}
\usepackage{ragged2e}
\usepackage{multirow}
\usepackage{helvet}
\usepackage{booktabs}
\usepackage{colortbl}
\usepackage{amsmath}
\usepackage{comment}
\usepackage{cmbright}

\usepackage{hyperref}
\hypersetup{
    colorlinks=true,
    linkcolor=sared,
    filecolor=magenta, 
    citecolor=sared,
    urlcolor=red,
}

\usepackage{blindtext}

\makeatletter
\newenvironment{figurehere}
{\def\@captype{figure}}
{}
\makeatletter

\usepackage{ccaption}
\usepackage{siunitx}
\usepackage{fixltx2e}
\newenvironment{sciabstract}{%
\begin{quote} \bf}
{\end{quote}}

\definecolor{sared}{rgb}{0.69, 0,0}

\newcounter{lastnote}

\title{\textbf{In-Situ Encryption of Single-Transistor Nonvolatile Memories without Density Loss}}

\author
{Sanwar Ahmed Ovy$^{1\dag}$, Jiahui Duan$^{2\dag}$, Md Ashraful Islam Romel$^{1}$,  \\Franz Müller$^{3}$, Thomas Kämpfe$^{3, 4}$, Kai Ni$^{2}$, Sumitha George$^{1*}$
\\
\\
\normalsize{$^{1}$North Dakota State University, Fargo, USA;}\\
\normalsize{$^{2}$University of Notre Dame, Notre Dame, USA;}\\
\normalsize{$^{3}$Fraunhofer IPMS, Dresden, Germany;}\\
\normalsize{$^{4}$TU Braunschweig, Braunschweig, Germany;}\\
\normalsize{$^\dag$Equal contribution;}\\
\normalsize{$^\ast$To whom correspondence should be addressed; E-mail:}\\ \normalsize{sumitha.george@ndsu.edu,kni@nd.edu,thomas.kaempfe@ipms.fraunhofer.de}
}

\date{}

\begin{document} 

\maketitle 
\begin{sciabstract}

Non-volatile memories (NVMs) offer negligible leakage power consumption, high integration density, and data retention, but their non-volatility also raises the risk of data exposure. Conventional encryption techniques such as the Advanced Encryption Standard (AES) incur large area overheads and performance penalties, motivating lightweight XOR-based in-situ encryption schemes with low area and power requirements. This work proposes an ultra-dense single-transistor encrypted cell using ferroelectric FET (FeFET) devices, which, to our knowledge, is the first to eliminate the two-memory-devices-per-encrypted-cell requirement in XOR-based schemes, enabling encrypted memory arrays to maintain the same number of storage devices as unencrypted arrays. The key idea is an in-memory single-FeFET XOR scheme, where the ciphertext is encoded in the device’s threshold voltage and leverages the direction-dependent current flow of the FeFET for single-cycle decryption; eliminating complementary bit storage also removes the need for two write cycles, allowing faster encryption. We extend the approach to multi-level-cell (MLC) FeFETs to store multiple bits per transistor. We validate the proposed idea through both simulation and experimental evaluations. Our analysis on a 128×128-bit array shows 2× higher encryption/decryption throughput than prior FeFET work and 45.2×/14.12× improvement over AES, while application-level evaluations using neural-network benchmarks demonstrate average latency reductions of 50\% and 95\% compared to prior FeFET-based and AES-based schemes, respectively.

\end{sciabstract}
\justify
\section*{\textcolor{sared}{Introduction}}
\label{sec:introduction_n}

The rapid rise of artificial intelligence (AI) is driven by promising applications across science \cite{scientific_discovery_in_the_age_of_artificial_intelligence}, robotics \cite{AI_driven_IoT_in_robotics_A_review}, art \cite{testing_the_capability_of_ai_art_tools_for_urban_design}, mathematics \cite{advancing_mathematics_by_guiding_human_intuition_with_ai}, and autonomous systems \cite{the_cognitive_internet_of_vehicles_for_autonomous_driving}. This momentum has accelerated the development and deployment of deep neural networks (DNNs) for diverse tasks. With the proliferation of the Internet of Things (IoT), ever more data are generated, processed, communicated, and stored at the edge \cite{data_encryption_based_on_field_effect_transistors_and_memristors}. However, training high-performance AI models directly on edge devices remains challenging due to their limited computational resources and strict power constraints \cite{model_compression_and_hardware_acceleration_for_neural_networks_a_comprehensive_survey}. Supervised training requires massive amounts of private labeled data, substantial computational resources, and significant time—often weeks or months. As a result, models are typically trained on resource-rich cloud servers, and the resulting model parameters or weights are later transferred to edge devices via wireless networks or stored locally \cite{xor_cim_compute_in_memory_sram_architecture_with_embedded_xor_encryption}. Unfortunately, this exposes the model weights to security threats (Fig.~\ref{fig:motivation}a). If an adversary gains access to weights -via memory readout -the model can be cloned into counterfeit chips or reverse engineered with retraining \cite{secure_rram_a_40nm_16kb_compute_in_memory_macro_with_reconfigurability_sparsity_control_and_embedded_security,xor_cim_compute_in_memory_sram_architecture_with_embedded_xor_encryption}. Securing stored weights is therefore essential.
Since weights are typically stored in memory arrays, it is critical to mask or encrypt the contents of these arrays. Among conventional memories, SRAM is widely used because of its speed and ease of integration; however, it is volatile and requires at least six transistors per cell, resulting in lower density \cite{hardware_functional_obfuscation_with_ferroelectric_active_interconnects}. Consequently, non-volatile memories (NVMs) have emerged as attractive alternatives for storage due to their negligible leakage power and high integration density. NVMs are particularly suitable for area- and power-constrained environments such as edge devices \cite{a_survey_of_techniques_for_improving_security_of_non_volatile_memories}.


Irrespective of the underlying memory technology, encrypting the memory array with a secret key is critical to mitigate risks of hacking, information leakage, and identity fraud \cite{security_of_iot_systems_design_challenges_and_opportunities, physical_unclonable_functions, innovations_in_hardware_security_leveraging_fefet_technology_for_future_opportunities}. Among existing encryption standards, the Advanced Encryption Standard (AES) \cite{the_rijndael_block_cipher_aes_proposal_a_comparison_with_des} is the de facto choice for protecting data at rest and in transit (Fig.~\ref{fig:motivation}b). AES is a symmetric block cipher operating on 128‑bit blocks with keys of 128, 192, or 256 bits. Despite strong security and algorithmic efficiency, deploying AES broadly on resource‑constrained edge devices is costly in energy and area. DNN deployments may require multiple AES modules to meet throughput demands; each module adds nontrivial area and power, and external decryption introduces additional latency \cite{first_demonstration_of_unclonable_double_encryption_28nm_rram_based_compute_in_memory_macro_for_confidential_ai,machine_learning_assisted_side_channel_attack_countermeasure_and_its_application_on_a_28_nm_aes_circuit}. 
To mitigate performance overheads, prior work \cite{securing_emerging_nonvolatile_main_memory_with_fast_and_energy_efficient_aes_in_memory_implementation} proposed AIM (AES In-Memory), which integrates an in-memory AES engine to perform bulk encryption of data blocks in NVMs. AIM triggers encryption only when necessary and, by leveraging in-memory computation to reduce data movement, thus achieving high encryption efficiency. However, its bulk-encryption design limits support for fine-grained protection and still requires energy-consuming AES operations (e.g., XORs, shifts, and table lookups). In summary, existing AES-based schemes do not efficiently address NVM security without incurring non-trivial costs. Therefore, lightweight encryption schemes are needed to resolve the trade-off between encryption/decryption performance and overhead \cite{embedding_security_into_ferroelectric_fet_array_via_in_situ_memory_operation}.

In‑memory encryption and decryption present a lightweight security solution that leverages intrinsic memory‑array operations without introducing complex encryption or decryption circuitry overhead, as illustrated in Fig.~\ref{fig:motivation}c \cite{embedding_security_into_ferroelectric_fet_array_via_in_situ_memory_operation}. This approach helps improve both throughput and energy efficiency compared to AES‑based schemes. The core idea is to implement a bitwise XOR between a secret key and the plaintext/ciphertext using the array’s native read/write primitives. Ciphertext is written via standard memory writes, and the data remain indecipherable unless the correct key is applied during sensing. In‑situ XOR‑based schemes have been demonstrated across multiple technologies \cite{secure_rram_a_40nm_16kb_compute_in_memory_macro_with_reconfigurability_sparsity_control_and_embedded_security,in_situ_encrypted_nand_fefet_array_for_secure_storage_and_compute_in_memory,embedding_security_into_ferroelectric_fet_array_via_in_situ_memory_operation}. Many implementations rely on complementary storage, multi‑cycle operations, or both \cite{in_situ_encrypted_nand_fefet_array_for_secure_storage_and_compute_in_memory,embedding_security_into_ferroelectric_fet_array_via_in_situ_memory_operation}. For example, for SRAM, a conventional 6T cell has been augmented with dual wordlines to create the XOR cipher \cite{xor_cim_compute_in_memory_sram_architecture_with_embedded_xor_encryption}. For RRAM, 2T2R cells leveraging complementary conductance states enable lightweight XOR encryption \cite{secure_rram_a_40nm_16kb_compute_in_memory_macro_with_reconfigurability_sparsity_control_and_embedded_security}. To enhance encryption robustness, another 2T2R RRAM-based compute-in-memory (CIM) design employs separate encryption keys for the sign and magnitude of neural-network weights, thereby strengthening security at the cost of additional processing overhead \cite{first_demonstration_of_unclonable_double_encryption_28nm_rram_based_compute_in_memory_macro_for_confidential_ai}. The most compact solution so far has been an encrypted cell made of complementary memory devices, each based on a single transistor memory, for example, ferroelectric field effect transistor (FeFET) \cite{embedding_security_into_ferroelectric_fet_array_via_in_situ_memory_operation,imce_an_in_memory_computing_and_encrypting_hardware_architecture_for_robust_edge_security,in_situ_encrypted_nand_fefet_array_for_secure_storage_and_compute_in_memory}. Among various candidates for implementing these encrypted memory architectures, ferroelectric HfO\textsubscript{2}-based FeFETs stand out for their scalability, CMOS compatibility, and energy efficiency, motivating the focus of this work on FeFET memory.

\begin{figurehere}
\centering
\includegraphics[width=0.95\textwidth,keepaspectratio]{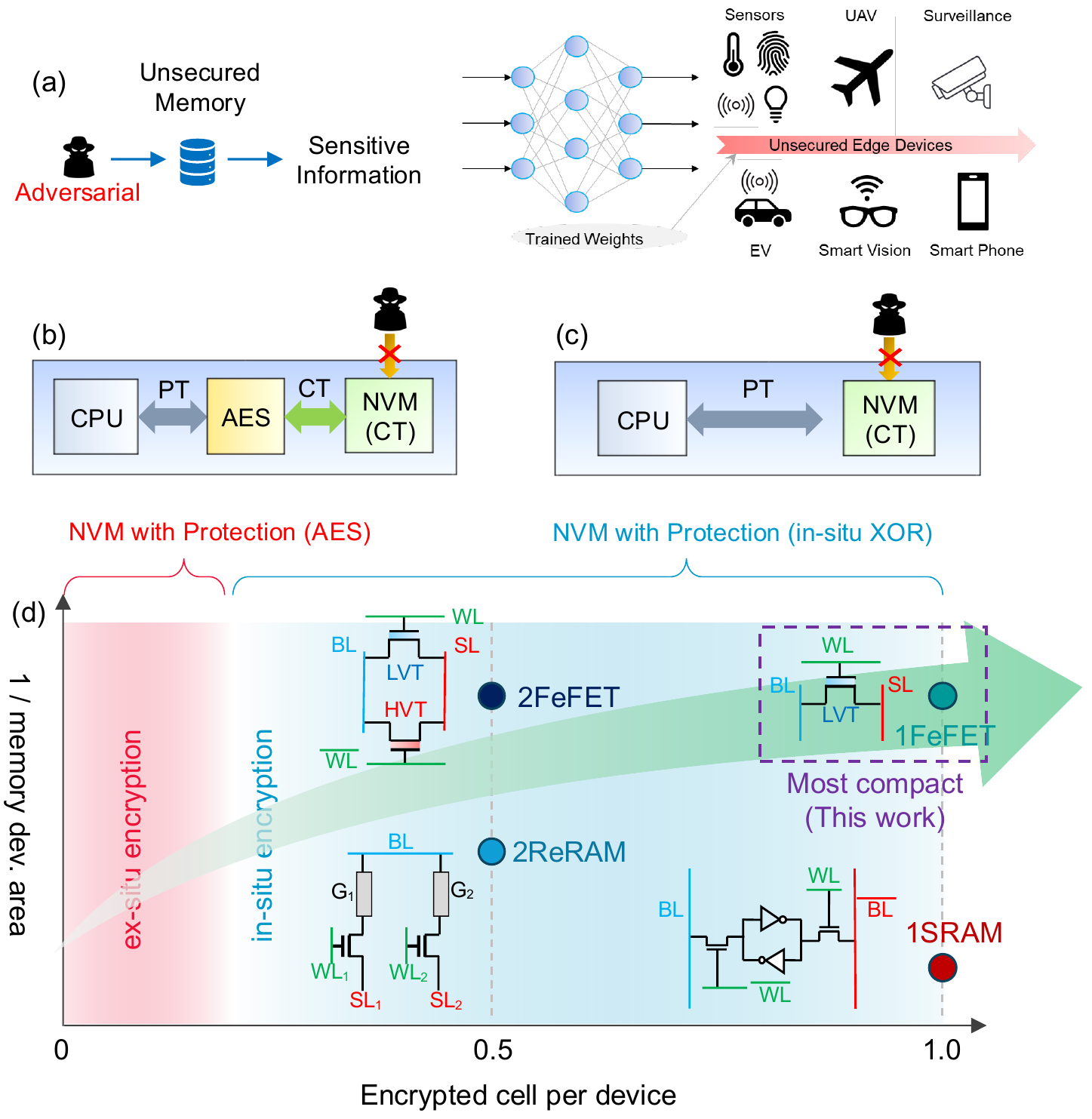}

\caption{\textbf{An overview of lightweight encrypted memories, and our proposed ultra‑compact single‑cell encrypted‑memory design.} (a) Need for NVM encryption. Unprotected non‑volatile memory (NVM) can leak sensitive data; in typical workflows. AI models are trained in the cloud and their weights are deployed to edge devices with memories vulnerable to memory‑theft attacks. If weights are stored as plaintext in NVM, an adversary with physical access can read them and clone the design. (b) Securing data with AES. Storing encrypted weights prevents disclosure, but a standalone AES (Advanced Encryption Standard) engine incurs significant latency, area, and energy overheads in resource‑constrained edge devices. (c) Lightweight in‑situ encryption/decryption. In‑memory (in‑situ) encryption removes the external encryption engine and its associated overheads. (d) Prior XOR‑cipher memories vs. proposed 1T XOR cipher. Prior RRAM/SRAM and FeFET XOR‑cipher designs require multiple devices per encrypted bit; in contrast, our work realizes a single‑cell, one‑transistor (1T) FeFET‑based XOR cipher that performs single‑cycle in‑memory XOR, enabling ultra‑high‑density secure memories with single-cycle decryption.}

\label{fig:motivation}
\end{figurehere}

FeFET memory cells have demonstrated superior performance relative to other emerging non-volatile memories \cite{Ferroelectric_field-effect_transistors_based_on_HfO2:_a_review,Ferroelectric_field_effect_transistors:_Progress_and_perspective, First_demonstration_of_in-memory_computing_crossbar_using_multi-level_Cell_FeFET}. In a FeFET, the ferroelectric layer is integrated as the gate dielectric of a MOSFET, where information is stored in the direction of the ferroelectric polarization, which can be switched by an applied electric field \cite{a_comprehensive_model_for_ferroelectric_fet_capturing_the_key_behaviors_scalability_variation_stochasticity_and_accumulation}. By configuring the polarization to point toward either the semiconductor channel or the gate electrode, the device can be programmed into a low-V\textsubscript{TH} or high-V\textsubscript{TH} state, respectively \cite{hardware_functional_obfuscation_with_ferroelectric_active_interconnects}.
This single-transistor memory architecture eliminates the need for a separate access device and, with decoupled read/write paths, improves operational stability compared to many alternatives. To further increase bit density, partial polarization switching in thin ferroelectric films can be leveraged to realize multiple threshold-voltage (V\textsubscript{TH}) levels, enabling multi-level-cell (MLC) operation in FeFETs \cite{Multibit_Ferroelectric_FET_Based_on_Nonidentical_Double_HfZrO2_for_High_Density_Nonvolatile_Memory, A_Novel_Ferroelectric_Superlattice_Based_Multi-Level_Cell_Non-Volatile_Memory}. FeFETs have been demonstrated on advanced transistor platforms—including 22-nm fully-depleted silicon-on-insulator (FDSOI) \cite{A_FeFET_based_super_low_power_ultra_fast_embedded_NVM_technology_for_22nm_FDSOI_and_beyond}, FinFET \cite{High_Speed_and_Large_Memory_Window_Ferroelectric_HfZrO₂_FinFET_for_High-Density_Nonvolatile_Memory}, and gate-all-around (GAA) devices \cite{Ferroelectric_Vertical_Gate_All_Around_Field_Effect_Transistors_With_High_Speed_High_Density_and_Large_Memory_Window}—highlighting strong scaling prospects. Reported properties include low read latency ($\sim$1 ns), current-source-like drive capability, ultra-low write energy ($<$ 1 fJ) with short write pulses ($\sim$10ns) \cite{First_demonstration_of_in-memory_computing_crossbar_using_multi-level_Cell_FeFET}, and excellent retention (up to 10 years) \cite{superior_qlc_retention_10_years_85c_and_record_memory_window_12_2_v_by_gate_stack_engineering_in_ferroelectric_fet_from_mifis_to_mikfis}, making FeFETs attractive for both memory and in-memory computing (IMC).

Prior lightweight implementations of FeFET-based XOR security schemes\cite{embedding_security_into_ferroelectric_fet_array_via_in_situ_memory_operation,imce_an_in_memory_computing_and_encrypting_hardware_architecture_for_robust_edge_security,in_situ_encrypted_nand_fefet_array_for_secure_storage_and_compute_in_memory} employ pairs of complementary FeFET devices to implement the XOR cipher. In these designs, each ciphertext bit is encoded by two FeFETs with opposite polarization states, storing one encrypted bit across a complementary device pair and thereby doubling the required memory-array capacity relative to conventional unencrypted storage. Xu et al. \cite{embedding_security_into_ferroelectric_fet_array_via_in_situ_memory_operation}  demonstrate an AND-type memory array using two FeFETs per unit cell, where the transistors are connected in parallel at their source–drain terminals and designed with complementary thresholds (Fig.~\ref{fig:implementation}b). Because decryption keys are applied through the shared wordline, decrypting a fine-grained key-protected block requires two wordline activations (two cycles) per row.  Shao et al \cite{imce_an_in_memory_computing_and_encrypting_hardware_architecture_for_robust_edge_security} extends complementary-cipher storage to support in-memory multiply–accumulate (MAC) on encrypted weights. Zijian et al. \cite{in_situ_encrypted_nand_fefet_array_for_secure_storage_and_compute_in_memory} adopt the XOR-cipher technique for vertical NAND FeFETs in which the ciphertext is stored as the configuration of two cells, and decryption is performed by applying key-dependent read wordline biases to two pages simultaneously—likewise doubling storage capacity and requiring two read cycles for fine-grained key-encrypted block decryption.

So far, all reported in-situ XOR schemes incur a 100\% area penalty to achieve security. The most desirable approach would therefore be to embed security directly within the memory array without sacrificing effective memory density—that is, to realize an in-situ XOR operation using a single memory device. Although several studies have explored in-memory XOR logic using a single memory element, these efforts have not been pursued in the encryption context \cite{Ferroelectric_FET-based_In-memory_1_Transistor_XOR_and_Majority_gate_for_Compact_and_Energy_Efficient_Hyperdimensional_Computing_Accelerator,Computing_in_memory_with_FeFETs}.
For example, Reis et al. \cite{Computing_in_memory_with_FeFETs} introduced a word-level in-memory XOR architecture in which both inputs are stored using a 2T+1FeFET configuration per memory bit, and the XOR operation is resolved through complex sense-amplifier circuitry. Although this approach employs only one memory device, each cell effectively involves three devices, which remains far from the goal of an ultimate single-device design.
Separately, Chakraborty et al. \cite{Ferroelectric_FET-based_In-memory_1_Transistor_XOR_and_Majority_gate_for_Compact_and_Energy_Efficient_Hyperdimensional_Computing_Accelerator} demonstrated a single-transistor XOR gate based on an FDSOI FeFET, utilizing a fundamentally different mechanism. In their design, two inputs are applied sequentially across the device terminals (gate, source, and drain), and a drain-erase–assisted programming sequence encodes the XOR result into the final V\textsubscript{TH} state. However, realizing this function requires multiple write/inhibit steps to traverse successive threshold levels, leading to significant latency and energy overheads. 

Given these challenges, we propose, for the first time, an ultra-compact in-situ XOR operation using a single memory device, enabling encryption without any density penalty. As illustrated in Fig.~\ref{fig:motivation}d, a single FeFET functions as an encrypted cell by directly storing the ciphertext. The density advantage is achieved through a novel in-memory XOR-based decryption mechanism that operates without complementary storage. Leveraging the three-terminal nature of FeFETs, the proposed compact cryptographic memory primitive employs a source/drain current-direction key-decoding scheme to enable single-cycle decryption, while the FeFET’s V\textsubscript{TH} states directly store encrypted data. This approach eliminates the need for auxiliary complementary bit storage required in prior lightweight XOR-based schemes, paving the way toward an extremely compact memory architecture that seamlessly integrates storage and security functionalities within a single device.
Moreover, since the decryption key is no longer distributed through a shared horizontal wordline, the design supports fine-grained, device-level decryption within a single cycle. Finally, we demonstrate that the scheme naturally extends to exploit the multilevel-cell (MLC) capability of FeFETs, enabling multiple encrypted bits to be stored per device—further reducing the encrypted array footprint and advancing bit-density scaling. Also, our proposed scheme can be extended to other 1T-1R memories as well. 

\justify
\section*{\textcolor{sared}{Working Principles and Validation}}
\label{sec:data_n}

\begin{flushleft} 
\textbf{\large Overview of the proposed in-memory encryption/decryption scheme}
\end{flushleft}
\vspace{-2ex}

Fig.~\ref{fig:implementation}(a) provides an overview of the proposed in situ XOR-based encryption/decryption scheme integrated with the FeFET memory array and its associated peripheral circuitry. In plaintext (unencrypted) operation, a single FeFET cell stores one bit (0/1); accordingly, each cell is programmed to a low- or high-threshold voltage state (LVT/HVT) depending on the data value. In our encrypted design, each cell is protected with a 1-bit key without any storage overhead, preserving ultra-dense memory (one encrypted bit per cell). For 1-bit encrypted storage, each cell is programmed to either HVT or LVT, with the FeFET driven to the desired state by applying write pulses of $\pm V_W$( Fig.~\ref{fig:writing implementation}). In our scheme, keys do not need to be shared across a row, enabling fine-grained per-cell encryption; successive cells can be encrypted with key 0 or key 1 without performance degradation.

\begin{figurehere}
\centering
\includegraphics[width=0.85\textwidth,keepaspectratio]{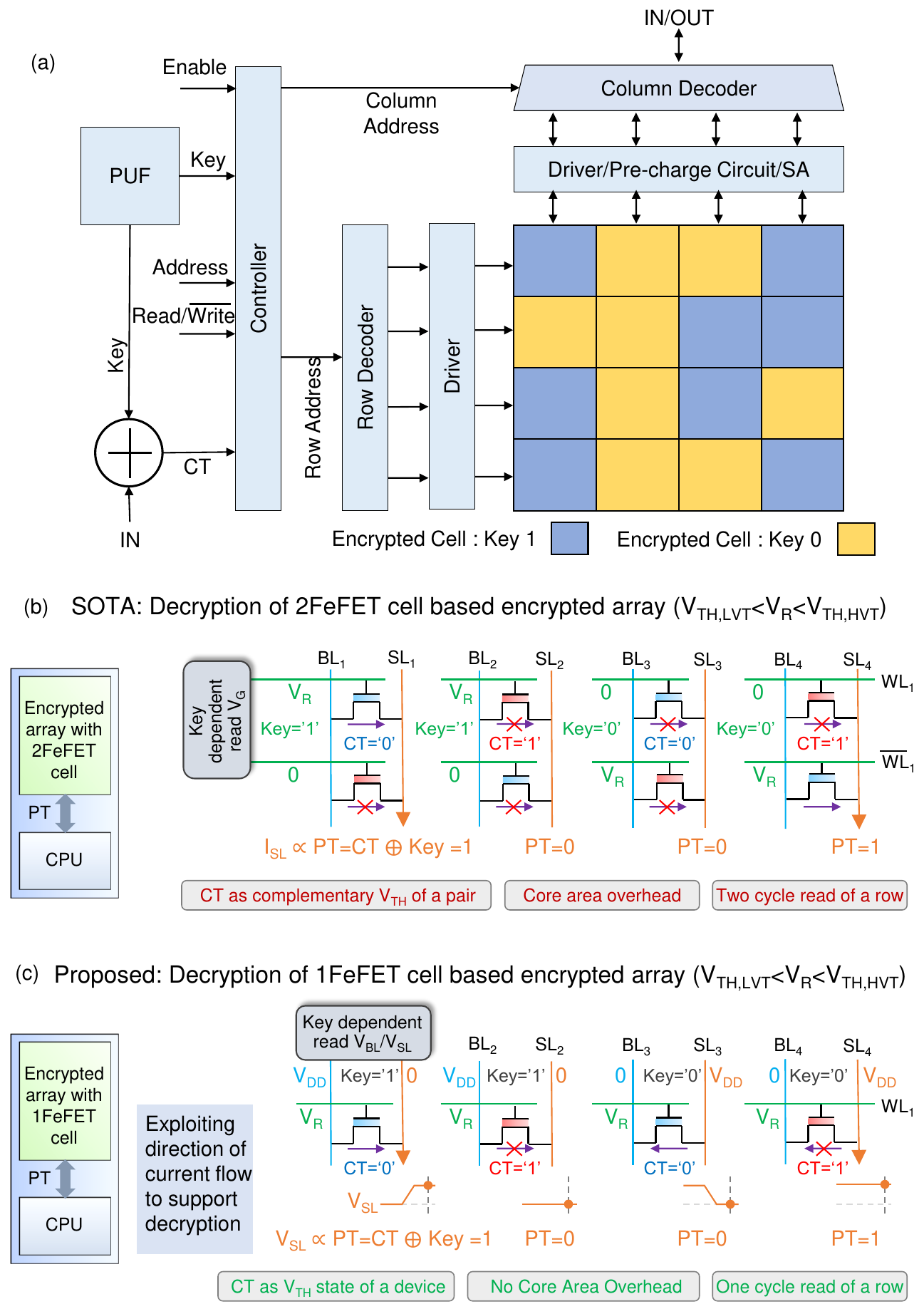}

\caption{\textbf{1T XOR-based in-memory encryption/decryption.}
(a) Architectural overview of the proposed lightweight in-memory XOR scheme.
(b) Prior FeFET-based state-of-the-art XOR designs use a two-transistor (2T) cell that stores complementary bits. This structure requires two decryption (read) cycles, with the key bit (0/1) applied on the same row \cite{embedding_security_into_ferroelectric_fet_array_via_in_situ_memory_operation}.
(c) Proposed 1T FeFET XOR scheme: the ciphertext is encoded in a single transistor by programming the device threshold state (HVT/LVT). Decryption is performed by applying the key as a source–drain (S/D) bias; the plaintext bit is obtained by sensing the source node, which resolves high or low depending on the S/D current direction. Different keys (0/1) can be applied in parallel across a row within the same cycle.}
\label{fig:implementation}
\end{figurehere}

Prior in‑situ FeFET encryption~\cite{embedding_security_into_ferroelectric_fet_array_via_in_situ_memory_operation} required two transistors per ciphertext bit (Fig.~\ref{fig:implementation}(b)). Each ciphertext bit was encoded in a complementary FeFET pair where one FeFET is programmed to HVT and the other one programmed to LVT or vice versa. Decryption recovered the plaintext by sensing the sense‑line current \(I_{\mathrm{SL}}\) while applying complementary read voltages \((V_\mathrm{R}, 0)\) to the two gates. The key bit selects which transistor’s gate receives \(V_\mathrm{R}\) and which receives \(0\); depending on the HVT/LVT assignment and the applied key, the sensed current resolves to a logical high or low, revealing the plaintext bit. Using a two‑FeFET (2T/bit) unit cell doubles the encrypted area and—when bits within a row use different keys—requires two read cycles per row because the row’s gate lines are shared and the key pair \((V_\mathrm{R}, 0)\) is broadcast per row, increasing decryption latency.

In contrast, our proposed in-memory encryption stores each encrypted bit in a single FeFET cell, preserving an ultra-dense encrypted array (1T/bit). During encryption, the ciphertext \(CT\) is generated by XORing the plaintext \(PT\) with the key \(Key\): \(CT = PT \oplus Key\). When \(Key=0\), \(CT=PT\) (plaintext \(0/1\) maps to ciphertext \(0/1\)); when \(Key=1\), \(CT=\overline{PT}\) (plaintext \(0/1\) maps to ciphertext \(1/0\)). We encode \(CT=0\) as a low-threshold (LVT) state and \(CT=1\) as a high-threshold (HVT) state in the FeFET cell, as illustrated in Fig.~\ref{fig:implementation}(c).
Decryption recovers the plaintext by XORing the ciphertext with the key, $PT = CT \oplus Key$. 
We realize this XOR in a \emph{single‑transistor, single‑cycle} read by treating the FeFET’s V\textsubscript{TH} state as one operand (i.e., CT) and the BL/SL polarity as the other (i.e., decryption key). 
The key is encoded in the drain‑/source‑line biases. 
For $Key=1$, initialize $V_{SL}=0$ (GND) and bias $V_{BL}=V_{DD}$; for $Key=0$, precharge $V_{SL}=V_{DD}$ and tie $V_{BL}=0$ (GND).
A read pulse $V_R$ is applied on the word line (WL) with $LVT<V_R<HVT$, so an LVT cell ($CT=0$) turns on while an HVT cell ($CT=1$) remains off. 
The plaintext is the final voltage on the source line (SL) after the read: if SL resolves to $V_{DD}$, then $PT=1$; if SL resolves to GND, then $PT=0$. 
Intuitively, when $CT=0$ (LVT), the channel conducts and SL is driven toward BL, yielding $PT=BL=Key$; when $CT=1$ (HVT), the channel is off and SL holds its precharge, yielding $PT=SL_{\text{init}}=\overline{Key}$. 
Thus the operation implements $PT = CT \oplus Key$ in one cycle.

Fig.~\ref{fig:implementation}(c)  summarizes the four decryption combinations of a 1-bit ciphertext and 1-bit key. For \(Key=1,\,CT=0\) (i.e., \(Key \oplus CT = 1 \oplus 0\)), the ciphertext maps to a LVT FeFET. A key of ‘1’ sets \(SL=0\) (GND) and \(BL=1\) (\(V_{\mathrm{DD}}\)). When the wordline (WL) is driven by the read voltage \(V_R\), chosen such that \(\mathrm{LVT}<V_R<\mathrm{HVT}\), the SL potential rises toward \(V_{\mathrm{DD}}\), yielding \(PT=1\). For \(Key=1,\,CT=1\) (\(1 \oplus 1\)), the ciphertext corresponds to a high-threshold (HVT) FeFET. With \(SL=0\) and \(BL=1\), the channel remains off at \(V_R\), the SL remains at GND, and \(PT=0\).
For \(Key=0,\,CT=0\) (\(0 \oplus 0\)), the ciphertext corresponds to LVT. The key ‘0’ precharges \(SL=1\) (VDD) and sets \(BL=0\) (GND). When WL is pulsed to \(V_R\), the device conducts, pulling SL down to GND and producing \(PT=0\).For \(Key=0,\,CT=1\) (\(0 \oplus 1\)), the ciphertext corresponds to HVT. With \(SL=1\) and \(BL=0\), the device remains off at \(V_R\); SL stays near \(V_{\mathrm{DD}}\), yielding \(PT=1\).These outcomes are consistent with \(PT = CT \oplus Key\).  Also note that, because the decryption keys are applied to the vertically routed SL and BL lines in the memory, keys do not need to be shared across a row(Refer Fig. \ref {fig:array_read}). This enables fine-grained per-cell decryption: cells in the same row can use key~0 or key~1 independently, allowing row-wide decryption in a single cycle. Simulation verification is given in Fig.~\ref{fig:Cadence Waveform}. Threshold variation analysis in Fig.~\ref{fig:monte_carlo} shows that the PT0/PT1 states are distinguishable with the proposed scheme.

Fig. \ref{fig:implementation}(a) represents an overview of the proposed in situ XOR-based encryption / decryption scheme with FeFET memory array. We integrate a PUF-derived key \cite{Exploiting_FeFET_Switching_Stochasticity_for_Low-Power_Reconfigurable_Physical_Unclonable_Function} with the memory array to generate unique encryption keys for each device. For 1-bit encrypted storage, each cell is programmed to either a high- or low-threshold state (HVT or LVT) through coordinated control of the row decoder, column decoder, and driver circuits. Instead of using a shared key across a row, a fine-granularity encryption scheme is implemented. The ciphertext bit is generated by XORing the input plaintext with the key derived from physically unclonable function (PUF). During decryption, appropriate biases are applied through the wordline (WL), bitline (BL), and sourceline (SL) using the decoders, drivers and pre-charge circuitry, The sense amplifier (SA) detects the SL voltage to decrypt the binary plaintext.
Unique encryption keys for the memory array can be realized by deriving them from a PUF implemented in the memory array itself~\cite{Exploiting_FeFET_Switching_Stochasticity_for_Low-Power_Reconfigurable_Physical_Unclonable_Function}. A PUF is a hardware-security primitive that exploits innate physical variations of a device~\cite{Physical_Unclonable_Functions_and_Applications_A_Tutorial} to generate unique and non-replicable keys. NVM elements are particularly well suited for PUF implementation, since their internal operating mechanisms naturally produce random responses even under identical external stimuli. For example, FeFET-based NVMs exhibit stochastic polarization switching under identical programming pulses, which results in random variations in drain current~\cite{innovations_in_hardware_security_leveraging_fefet_technology_for_future_opportunities,Demonstration_of_high-reconfigurability_and_low-power_strong_physical_unclonable_function_empowered_by_FeFET_cycle-to-cycle_variation_and_charge-domain_computing}. Such stochastic switching behavior of the ferroelectric layer inherently yields distinct keys for every device. FeFET-based PUFs have been implemented in both single- and dual-transistor configurations~\cite{Exploiting_FeFET_Switching_Stochasticity_for_Low-Power_Reconfigurable_Physical_Unclonable_Function}. To reduce key volume and simplify key management, block-cipher-based encryption can be adopted instead of fine-grained per-cell encryption. In this scheme, each memory block is encrypted using a unique  key, ensuring that different blocks are still protected by distinct keys.

\begin{flushleft} 
\textbf{\large Functional verification}
\end{flushleft}

Next we show the experimental validation of the proposed encryption/decryption scheme shown in Fig.\ref{fig:implementation} using a FeFET AND array implemented on GlobalFoundries' 28nm high-$\kappa$ metal gate (HKMG) platform \cite{trentzsch201628nm}. In this work, an 8x6 FeFET AND array is adopted for proof of concept, as shown in Fig.\ref{fig:fabrication}(a). On-wafer probe card testing system is adopted for the array testing, as shown in Fig.\ref{fig:Experimental Setup}. The top-view optical image of the array shows the needles to contact pads and a total of 25 pads are used for this array. Detailed device information and processes can be found in \cite{trentzsch201628nm}. The cross-sectional transmission electron microscopy (TEM) of the FeFET is shown in Fig.\ref{fig:fabrication}(b), where the gate stack features a 8nm thick doped HfO\textsubscript{2} as the dielectric (per Fig.\ref{fig:fabrication}(c)), rather than the high-$\kappa$ dielectric in the logic device. Except for the gate dielectric difference, the FeFETs share the same processing as the logic device, and thus can be integrated side-by-side with logic device, making it versatile for various applications.

\begin{figurehere}
\centering
\includegraphics[width=0.85\linewidth]{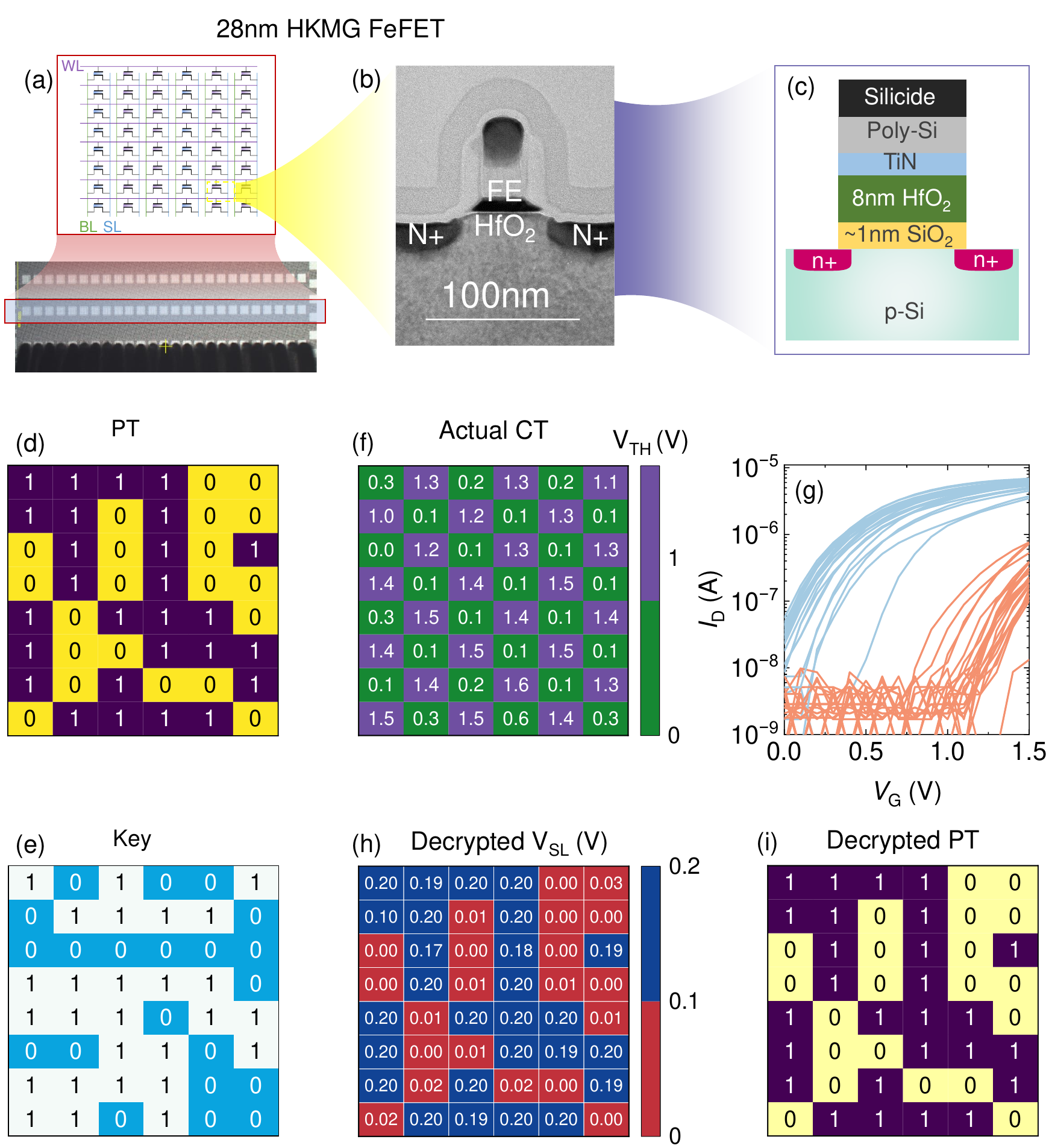}
\caption{\textbf{Experimental validation of in-situ encryption/decryption with SLC FeFET array.} (a) SEM top-view of 8x6 FeFET AND array for encryption and decryption demonstration. (b) The TEM cross-section and (c) schematics of the FeFET. (d) Original SLC plaintext for the 8×6 FeFET AND array. (f) The Vths map representing ciphertext and (g) corresponding transfer curves are measured, which reflects the encryption results using the random generated (e) key. Using the same correct key, (h) the decrypted SL voltage map represents the correct (i) decrypted PT, which has 100\% accuracy rate.}
\label{fig:fabrication}
\end{figurehere}

In the experimental demonstration, a checkerboard pattern of ciphertext is used. Figure~\ref{fig:fabrication}(d) illustrates the plaintext matrix representing  data input. Each cell corresponds to a binary bit, where yellow and purple cells denote logic ‘0’ and ‘1’, respectively. Figure~\ref{fig:fabrication}(e) shows the randomly generated key matrix, which is XOR-combined with the plaintext data to produce the checkerboard ciphertext distribution. In Fig.~\ref{fig:fabrication}(f), the ciphertext is mapped to V\textsubscript{TH} states of FeFETs. Values in the range of 0–0.6~V represent LVT states corresponding to CT ‘0’, whereas values in the range of 1.0–1.6~V represent HVT states corresponding to CT ‘1’. Figure~\ref{fig:fabrication}(g) presents the I\textsubscript{D}-V\textsubscript{G} characteristics of all FeETs in the array, with blue curves representing CT ‘0’ (LVT) and red curves representing CT ‘1’ (HVT). It shows clearly distinguished states. Then decryption operation, according to Fig.\ref{fig:implementation}(c), is executed. Key-dependent BL/SL voltages and read gate voltages are applied and corresponding SL voltages are measured. The resulting SL voltages are shown in Fig.~\ref{fig:fabrication}(h). SL voltages within the red highlighted boxes range from 0 to 0.1~V; interpreting 0–0.1~V as logic low  yields a plaintext value of ‘0’ for these cells. Similarly, SL voltages in the 0.1–0.2~V range are interpreted as logic high, highlighted in blue, and correspond to plaintext ‘1’. The binary mapping of these $V_{\mathrm{SL}}$ values is provided in Fig.~\ref{fig:fabrication}(i) as the decrypted output. At the single-cell (SLC) level, the decrypted output achieves 100\% accuracy relative to the original plaintext (PT) pattern in Fig.~\ref{fig:fabrication}(d).

\begin{flushleft}
{\large \textbf{MLC encryption decryption}\par}

\justify



The proposed scheme can also be extended to multi-level-cell (MLC) operation, enabling higher storage density while maintaining the same number of storage devices. Partial polarization states in the ferroelectric layer of FeFETs give rise to multiple distinct, nonvolatile V\textsubscript{TH} levels, which can be exploited to realize MLC storage. In this section, the 2-bit MLC encryption/decryption extension of the scheme is demonstrated. A 2-bit plaintext ($PT_{\text{MSB}}$,$PT_{\text{LSB}}$) is XORed with a 2-bit encryption key ($Key_{\text{MSB}}$,$Key_{\text{LSB}}$) to generate a 2-bit ciphertext($CT_{\text{MSB}}$,$CT_{\text{LSB}}$), where $
    CT_{\text{MSB}} = Key_{\text{MSB}} \oplus PT_{\text{MSB}};\quad
    CT_{\text{LSB}} = Key_{\text{LSB}} \oplus PT_{\text{LSB}};
$ The resulting 2-bit ciphertext values \(\{00, 01, 10, 11\}\) are encoded in a  single FeFET using four V\textsubscript{TH} states , as illustrated in Fig.~\ref{fig:MLC}(a). Figure~\ref{fig:MLC}(b) shows the \(I_D\text{--}V_G\) characteristics of FeFETs with four distinct V\textsubscript{TH} states. The lowest threshold voltage \(V_{\text{TH},00}\) is mapped to CT `00', the second-lowest \(V_{\text{TH},01}\) to CT `01', the second-highest \(V_{\text{TH},10}\) to CT `10', and the highest \(V_{\text{TH},11}\) to CT `11'. These mapped threshold voltages satisfy
\begin{equation}
    V_{\text{TH},00} < V_{\text{TH},01} < V_{\text{TH},10} < V_{\text{TH},11}.
\end{equation}




For 2-bit decryption, the plaintext is decoded from the sequence of sense-line (SL) outputs obtained during three FeFET read cycles. In these three cycles, the word line (WL) is biased sequentially with the read voltages \(V_{R2}\), \(V_{R1}\), and \(V_{R3}\), which are chosen such that
$V_{TH,00} < V_{R1} < V_{TH,01} < V_{R2} < V_{TH,10} < V_{R3} < V_{TH,11}$. Plaintext decryption is illustrated in Figs.~\ref{fig:MLC}(c) and (d). The most significant bit of the plaintext, \(PT_{\text{MSB}}\), is recovered in the first read cycle. The decryption of the least significant bit, \(PT_{\text{LSB}}\), depends on the ciphertext MSB, \(CT_{\text{MSB}}\), as shown in Fig.~\ref{fig:MLC}(c). If \(CT_{\text{MSB}} = 0\), then \(PT_{\text{LSB}}\) is obtained from the SL output in the second cycle; if \(CT_{\text{MSB}} = 1\), then \(PT_{\text{LSB}}\) is obtained from the SL output in the third cycle. The decryption key bits are applied as bias voltages on the BL/SL lines. For a key bit of 1, the BL is biased at \(V_{\text{DD}}\) while the SL is pre-charged to 0~V (BL/SL = 1/0), as illustrated in Fig.~\ref{fig:MLC}(d). For a key bit of 0, the BL is biased at 0~V and the SL is pre-charged to \(V_{\text{DD}}\) (BL/SL = 0/1).



\begin{figurehere}
\centering
\includegraphics[width=1\linewidth]{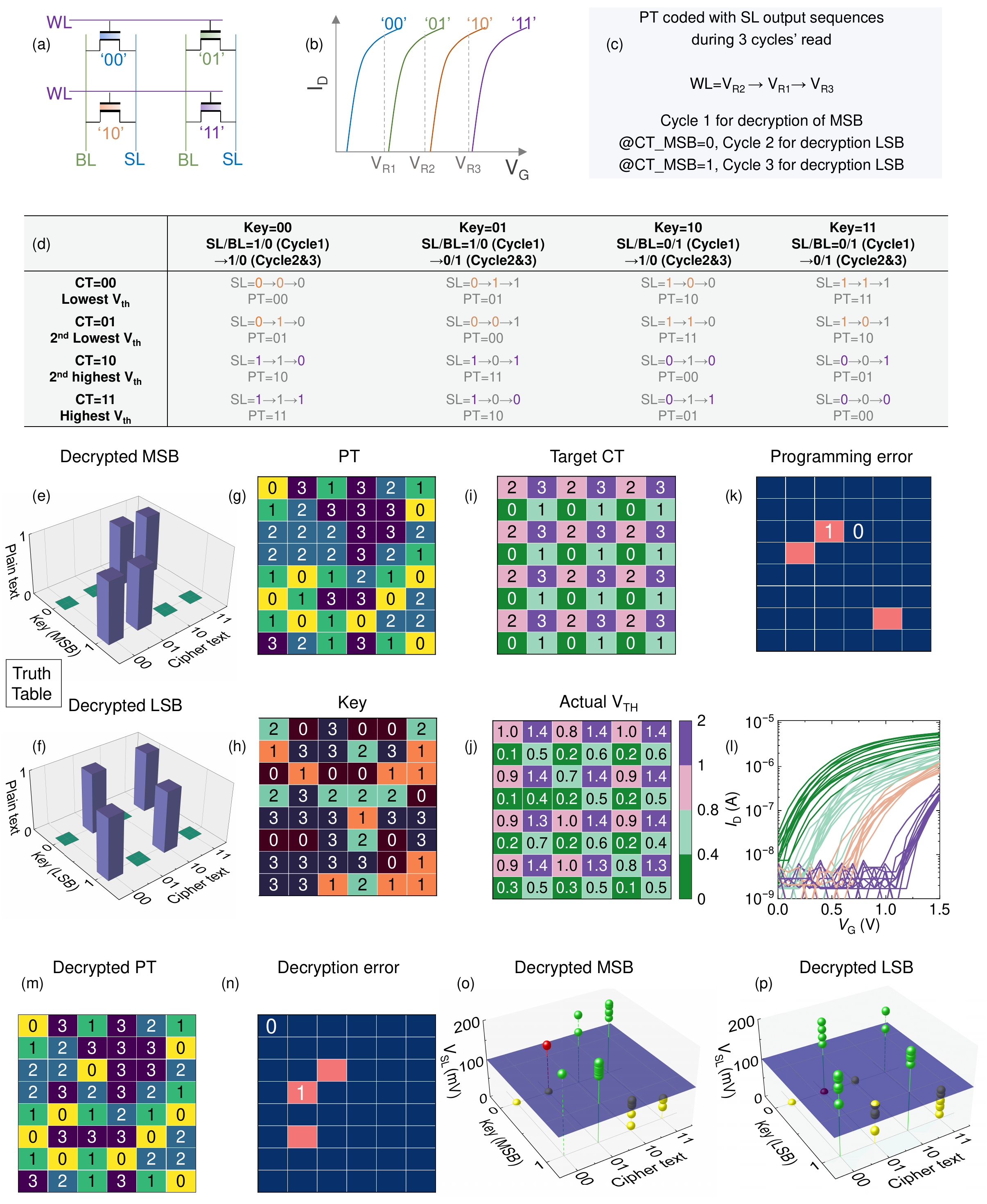}
\caption{\textbf{Experimental validation of in-situ encryption/decryption with MLC FeFET array.} (a) Schematics of four states FeFET AND array and the corresponding (b) transfer curves. (c) Three cycles are required to decrypt the MLC data, where VR2, VR1, VR3 are applied to WL sequentially. and (d) the expected SL voltage during each cycle. The truth table of decrypting (e) MSB and (f) LSB. (g) Original PT checkboard is used for demonstration. By using the (h) key, The (i) target CT is expected representing the XOR result of PT and Key. After encrypting the MLC data, the (j) Vths map is extracted, representing the actual CT and showing (k) 6.25\% error rate. The Vths map is extracted from the (l)transfer curves in this 8x6 array. By using the same key, the (m) PT is decrypted, showing also (n) 6.25\% error rate. The sensing voltage at SL during (o) MSB and (p)  LSB decryption step.}

\label{fig:MLC}
\end{figurehere}

Figure~\ref{fig:MLC}(d) summarizes all 16 possible combinations of the 2-bit ciphertext and 2-bit key. In the first read cycle, with \(V_{R2}\) applied to the WL, the most significant bit of the plaintext, \(PT_{\text{MSB}}\), is extracted. First consider the case \(Key_{\text{MSB}} = 1\), implemented by \(\text{BL/SL} = 1/0\) ($V_{BL}$ = \(V_{\text{DD}}\), $V_{SL}$ precharged to 0~V).For cells programmed to CT~`00' or CT~`01' (i.e., with \(V_{\text{TH},00}\) or \(V_{\text{TH},01}\)), \(V_{R2}\) is larger than the threshold voltage, the FeFET turns on, and \(V_{\text{SL}}\) is driven toward \(V_{\text{DD}}\). This SL level is decoded as \(PT_{\text{MSB}} = 1\). For cells programmed to CT~`10' or CT~`11' (with \(V_{\text{TH},10}\) or \(V_{\text{TH},11}\)), \(V_{R2}\) is below the threshold voltage, the FeFET remains off, and \(V_{\text{SL}}\) stays near 0~V, which is decoded as \(PT_{\text{MSB}} = 0\). Similarly, for \(Key_{\text{MSB}} = 0\), realized by \(\text{BL/SL} = 0/1\) ($V_{BL}$ = 0~V, $V_{SL}$ precharged to \(V_{\text{DD}}\)), cells with CT~`00' or CT~`01' (low-\(V_{\text{TH}}\) states) turn on and discharge SL from \(V_{\text{DD}}\) to 0~V, yielding \(PT_{\text{MSB}} = 0\), whereas cells with CT~`10' or CT~`11' (high-\(V_{\text{TH}}\) states) leave SL at \(V_{\text{DD}}\), yielding \(PT_{\text{MSB}} = 1\). Thus, in all cases the extracted \(PT_{\text{MSB}}\) satisfies $CT_{\text{MSB}} = Key_{\text{MSB}} \oplus PT_{\text{MSB}}$.

When $CT_{MSB}$ is 0 $(CT= `00'/'01') $, then we read out the LSB plaintext from SL voltage during second cycle with $V_{R1}$ on WL. If $Key_{LSB}=0 (BL/SL = 0/1)$, $V_{SL}$ would be driven to GND for CT '00', indicating the value for $PT_{LSB}$ to be 0. However, $V_{SL}$ stays at V\textsubscript{DD} for CT '01', indicating the value for $PT_{LSB}$ is 1. In contrast, if the $key_{LSB}$ is 1 (BL/SL=1/0), $V_{SL}$ is driven to V\textsubscript{DD} for CT `00', inferring  $PT_{LSB}$ to be 1. For CT `01', $V_{SL}$ stays at GND, which indicates the value for $PT_{LSB}$ is 0. However, if $CT_{MSB}$ is 1 $(CT= `10'/'11')$, then we read out the LSB plaintext from SL voltage during third cycle with $V_{R3}$ on WL. If $Key_{LSB}=0 (BL/SL = 0/1)$, $V_{SL}$ would be driven to GND for CT '10', indicating the value for $PT_{LSB}$ to be 0. However, $V_{SL}$ stays at V\textsubscript{DD} for CT '11', indicating the value for $PT_{LSB}$ is 1. In contrast, if the $key_{LSB}$ is 1 (BL/SL=1/0), $V_{SL}$ is driven to V\textsubscript{DD} for CT `10', inferring  $PT_{LSB}$ to be 1 while $V_{SL}$ stays at GND for CT of `11' and $Key_{LSB}$ of 1 (BL/SL=1/0), which indicates the value for $PT_{LSB}$ is 0. Simulation verification is shown in Fig.~\ref{fig:mlc_xor}.

Fig.~\ref{fig:MLC}(e) and (f) show the truth tables for the MSB and LSB of the plaintext, respectively. In Fig.~\ref{fig:MLC}(e), $CT$, $Key_{\mathrm{MSB}}$, and $PT_{\mathrm{MSB}}$ are mapped to the x-, y-, and z-axes, illustrating $PT_{\mathrm{MSB}} = CT_{\mathrm{MSB}} \oplus Key_{\mathrm{MSB}}$, while in Fig.~\ref{fig:MLC}(f), $CT$, $Key_{\mathrm{LSB}}$, and $PT_{\mathrm{LSB}}$ are mapped to the x-, y-, and z-axes, illustrating $PT_{\mathrm{LSB}} = CT_{\mathrm{LSB}} \oplus Key_{\mathrm{LSB}}$.
The same 8x6 FeFET array shown in Fig.\ref{fig:fabrication}(a) is adopted to validate the proposed MLC encryption/decryption scheme. Fig. \ref{fig:MLC}(g) represents the plaintext matrix where 2-bit plaintext '00', '01','10', '11' is represented as decimal numbers 0,1,2,3 respectively.  The value for the encryption key matrix is illustrated in Fig.\ref{fig:MLC}(h). Two-bit PT XORed with two-bit keys result into the target ciphertext distribution shown in Fig.\ref{fig:MLC}(i), shown in corresponding decimal numbers. This target CT matrix is programmed as V\textsubscript{TH} states of the FeFET, as shown in Fig.\ref{fig:MLC} (j). Based on the encoding rule shown in Fig.\ref{fig:MLC}(j), where V\textsubscript{TH} in the range of 0V-0.4V, 0.4V-0.8V, 0.8V-1V, and 1V-2V is mapped as ciphertext 0,1,2, and 3, respectively, there are some errors incurred during the encryption process, as shown in Fig.\ref{fig:MLC}(k). 
Comparing the target CT matrix with the actually programmed CT matrix yields an error rate of 3 out of 48 cells. The corresponding I\textsubscript{D}-V\textsubscript{G} characteristics for these 48 programmed cells are presented in Fig.~\ref{fig:MLC}(l). It clearly shows the overlap between different states. The encryption errors incurred by MLC programming are due to multiple reasons. First, the programming is done in a single-shot programming, where write-and-verify based iterative programming is not implemented. Second, tested FeFET has a limited memory window, around 1.2V, which is not easy to implement MLC. Typical flash based solid state drive has a window around 6V to realize MLC or TLC. It is expected that such encoding errors can be avoided in existing flash memory and also FeFET with a large memory window designed for storage applications \cite{zhao2024large}. In this work, we focus on validating the encryption/decryption principles and leaving the device optimization for the future research.

During decryption, voltages are applied according to the key matrix in Fig.~\ref{fig:MLC}(h), and the resulting SL voltages are mapped to the decrypted plaintext matrix in Fig.~\ref{fig:MLC}(m). A decryption error rate of 3 out of 48 cells is observed, as illustrated in Fig.~\ref{fig:MLC}(n). Out of the three decryption errors, two of them correspond to encryption failures, which can be fixed by write scheme on appropriate devices. Meanwhile, one encryption error disappears during decryption, and a new decryption error appears.
For the device at column 5, row 7, its actual V\textsubscript{TH} is around 0.8 V, which lies between data states 1 and 2. This overlap may have caused the encryption error. However, its current at $V_{R2}$ is close to that of data 1, so the error disappears during decryption. For the device at row 6, column 2, although the data are correctly encrypted, the current at $V_{R2}$ is quite low, leading to a decryption error.
Figures~\ref{fig:MLC}(o) and \ref{fig:MLC}(p) show the most significant bit (MSB) and least significant bit (LSB) components of the decrypted plaintext, respectively. Source-line voltages ($V_{SL}$) in the range 0–100~mV are classified as bit `0' and are plotted as yellow dots. Voltages in the range 100–200~mV are classified as bit `1' and are plotted as green dots. Red dots indicate decryption errors corresponding to the cells highlighted in Fig.~\ref{fig:MLC}(n). These results validate the encryption/decryption principles and the errors are expected to be eliminated with larger memory window and appropriate write scheme.

\section*{\textcolor{sared}{Benchmarking}}
\label{sec:benchmarking_n}
\justify
\vspace{-3ex}
We compare the proposed FeFET in-memory XOR encryption/decryption with (i) an AES-based scheme~\cite{A_923_Gbps/W_113-Cycle_2-Sbox_Energy-efficient_AES_Accelerator_in_28nm_CMOS} and (ii) a prior FeFET-based encryption design~\cite{embedding_security_into_ferroelectric_fet_array_via_in_situ_memory_operation}. The comparison spans area, latency, power, and throughput (Fig.~\ref{fig:seven NN comparison}). The analysis is performed for an encrypted memory array clocked at 25\,MHz (40\,ns cycle), aligned with previous AES work~\cite{A_923_Gbps/W_113-Cycle_2-Sbox_Energy-efficient_AES_Accelerator_in_28nm_CMOS}. We target an 128$\times$128 plaintext memory array to be encrypted. In this work, the encrypted FeFET memory uses a 128$\times$128 array, whereas the prior work uses a 256$\times$128 array. The design employs sixteen sense amplifiers (SAs) to decrypt one 128-bit row; each SA is time-multiplexed across eight columns (one column per cycle), so a full-row decrypt completes in eight cycles.
Our design achieves substantially lower area than prior works, as illustrated in Fig.~\ref{fig:seven NN comparison}(a). 
Earlier AES-based approaches report an area of $0.00309~\mathrm{mm}^2$ \cite{A_923_Gbps/W_113-Cycle_2-Sbox_Energy-efficient_AES_Accelerator_in_28nm_CMOS}, while both the prior FeFET design and our scheme incur negligible overhead comparing to the whole memory area cost as this scheme needs only some XOR gates used to generate ciphertext~\cite{embedding_security_into_ferroelectric_fet_array_via_in_situ_memory_operation}. Importantly, the core array in our scheme introduces \emph{zero} transistor overhead: the encrypted array uses the same number of devices as the plaintext array. In contrast, the prior FeFET in-memory encryption design doubles the device count (e.g., $256\times128$ FeFETs) relative to an equivalently sized plaintext array, whereas our encrypted array for a $128\times128$ configuration preserves the original device count~\cite{embedding_security_into_ferroelectric_fet_array_via_in_situ_memory_operation}. Based on layout analysis given in Fig.~\ref{fig:layout}, the area per 1-bit ciphertext in our design is reduced by 50\% compared to the previous FeFET-based approach~\cite{embedding_security_into_ferroelectric_fet_array_via_in_situ_memory_operation}, as shown in Fig.~\ref{fig:seven NN comparison}(d), resulting in significant reduction in the core array area.

\begin{figurehere}
	\centering
	\includegraphics [width=0.95\textwidth,keepaspectratio]{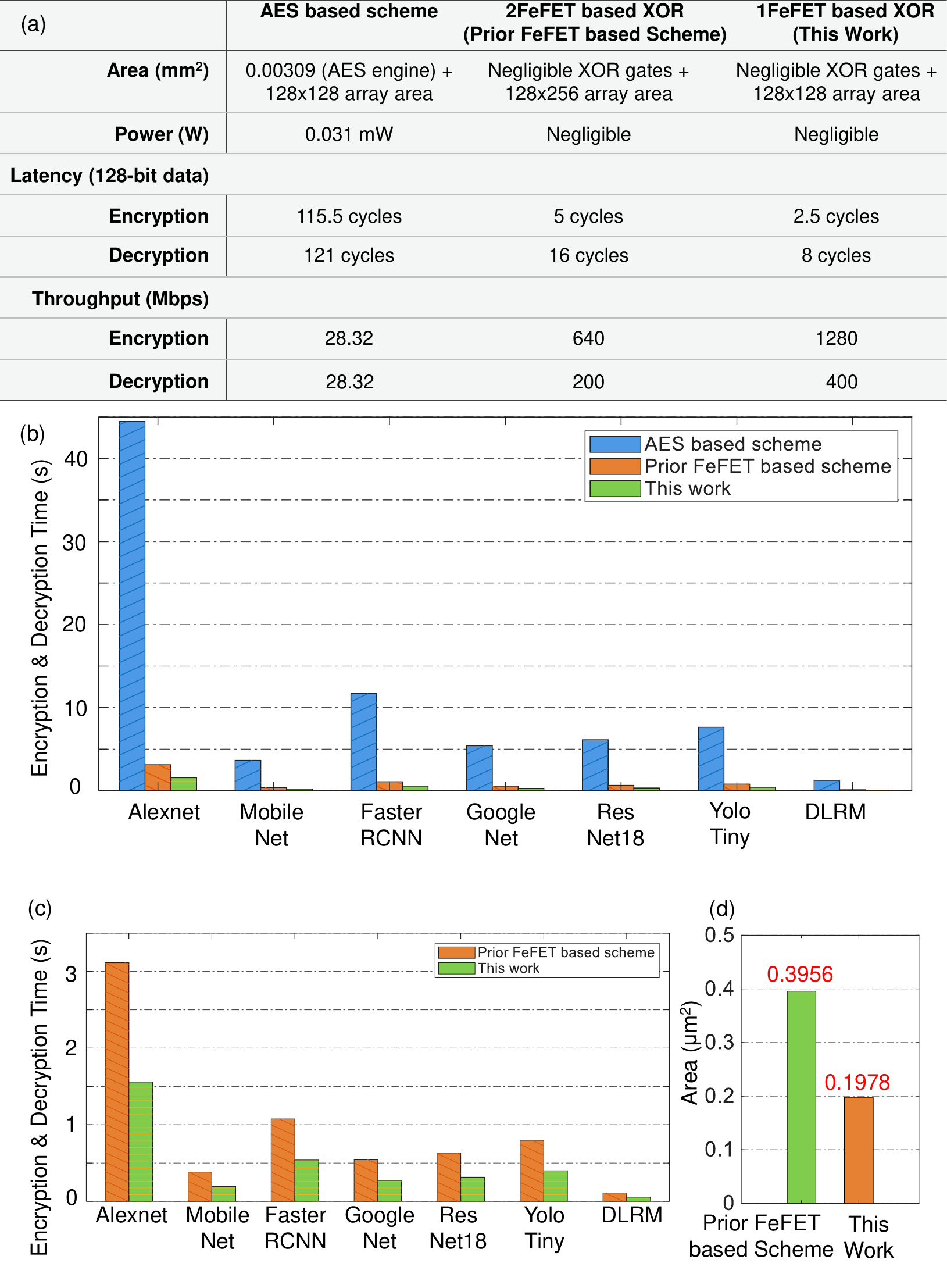}
    \caption{\textbf{Benchmarking of the proposed encryption/decryption scheme.} (a) Comparison of power, core area, latency, and throughput for encrypting a $128\times128$-bit plaintext among an AES accelerator~\cite{A_923_Gbps/W_113-Cycle_2-Sbox_Energy-efficient_AES_Accelerator_in_28nm_CMOS}, the prior FeFET-based design~\cite{embedding_security_into_ferroelectric_fet_array_via_in_situ_memory_operation}, and this work. Our design achieves encryption/decryption throughput that is $45.2\times/14.12\times$ higher than AES and $2\times/2\times$ higher than the prior FeFET design. (b) Performance comparison of AES (Weiwei Shan et al.~\cite{A_923_Gbps/W_113-Cycle_2-Sbox_Energy-efficient_AES_Accelerator_in_28nm_CMOS}), the prior FeFET scheme (Yixin Xu et al.~\cite{embedding_security_into_ferroelectric_fet_array_via_in_situ_memory_operation}), and this work across seven neural-network workloads, showing an average total latency reduction of $95\%$ relative to AES, and  (c) Total latency reduction of $50\%$ compared with the previous FeFET-based scheme. (d) Encrypted cell area reduction of $50\%$ compared with the prior FeFET work.}\label{fig:seven NN comparison}
\end{figurehere}

The proposed design demonstrates significant latency and throughput improvements over prior works. The array operates at 25 MHz (40 ns clock period), consistent with the AES work \cite{A_923_Gbps/W_113-Cycle_2-Sbox_Energy-efficient_AES_Accelerator_in_28nm_CMOS}. Our scheme uses a fixed 100 ns write window per 128-bit row, corresponding to 2.5 encryption cycles (100 ns / 40 ns = 2.5).
In our work, decryption takes one cycle per bit for both key values (0/1), enabling single-cycle, fine-grained decryption across the row. With sixteen sense amplifiers (SAs), the 128-bit word is partitioned into eight parts, so a full-row decrypt completes in eight cycles. Thus, our latency for a 128-bit row is (encryption, decryption) = (2.5, 8) cycles.
For 128-bit data using AES, the latency is (115.5, 121) cycles \cite{embedding_security_into_ferroelectric_fet_array_via_in_situ_memory_operation}, so our work achieves encryption/decryption latency improvements of 46.2×/15.13× over AES. Prior FeFET work uses 2 FeFETs per 1-bit ciphertext; therefore, it requires two write cycles for encryption. For decryption, two read cycles are also required due to wordline/key sharing (keys applied via wordlines) across rows, which prevents simultaneous fine-grained decryption. Its latency is (5, 16) cycles (encryption/decryption), i.e., 2×/2× higher than ours. One approach to further improve decryption latency/throughput would be to increase the number of SAs.
We also evaluate throughput at a 25 MHz clock frequency. Encrypting 128-bit data takes 2.5 cycles (100 ns), yielding 1,280 Mbps encryption throughput. With 16 SAs, decryption processes 16 bits per cycle (40 ns), yielding 400 Mbps decryption throughput. The (encryption, decryption) throughput for AES is (28.32 Mbps, 28.32 Mbps) \cite{A_923_Gbps/W_113-Cycle_2-Sbox_Energy-efficient_AES_Accelerator_in_28nm_CMOS}, and for the prior FeFET design is (640 Mbps, 200 Mbps). Note, in this analysis, we assume the case of different key bits applied in the same row. Therefore, our design achieves ~45.2×/14.12× higher throughput than AES and 2×/2× higher than the prior FeFET design.

Next we show how the array level performance is impacting the system level performance. To compare the impact of encryption/decryption latency across diverse neural-network workloads, we conduct a case study on AlexNet, MobileNet, Faster RCNN, GoogleNet, ResNet-18, YOLO-Tiny, and DLRM. We evaluate three schemes: a conventional AES baseline~\cite{A_923_Gbps/W_113-Cycle_2-Sbox_Energy-efficient_AES_Accelerator_in_28nm_CMOS}, a prior FeFET-based design~\cite{embedding_security_into_ferroelectric_fet_array_via_in_situ_memory_operation}, and our proposed approach. Experiments are performed with SCALE-Sim~\cite{SCALE-Sim:_Systolic_CNN_Accelerator_Simulator,A_systematic_methodology_for_characterizing_scalability_of_DNN_accelerators_using_SCALE-sim}, a widely used simulator for systolic CNN accelerators. All workloads execute on a TPU-like systolic array (weight-stationary dataflow). In our design, encrypted weights (ciphertext) reside in a 1T-FeFET memory array and are decrypted and provided as plaintext to the systolic array, which processes each layer under a weight-stationary dataflow. The computation outputs are then read, encrypted, and stored back to the FeFET array. Using SCALE-Sim, we obtain workload-specific reads and writes to the memory array. Our analysis shows that the encryption and decryption latency overhead is significantly reduced with our proposed in-situ encryption/decryption. Our scheme achieves an average latency reduction of 95\% compared to AES, as shown in Fig.~\ref{fig:seven NN comparison}(b). Relative to the prior FeFET-based design, our approach delivers an 50\% latency reduction as shown in Fig.~\ref{fig:seven NN comparison}(c), enabled by two architectural advantages: (i) one-bit-per-transistor (1T-FeFET) encoding and (ii) parallel application of decryption key bits (0/1) in a single read cycle. 
By contrast, the prior FeFET scheme uses two FeFETs per bit and requires two read cycles for fine-grained decryption, resulting in higher latency and reduced efficiency.

\section*{\textcolor{sared}{Discussion}}
\label{sec:discussion_n}
To summarize this work, we propose a XOR-based in-memory hardware-based encryption scheme. We use only one transistor for an unit cell, which makes the design ultra compact. We verify the functionality of the proposed design through software simulation and fabrication data. We also conduct a case-study on different neural network workloads to compare the performance benefit of the proposed design against state-of-the-art work. In ours, the individual keys are applied simultaneously which halves the decryption latency than the state-of-the-art work.
\justify
\section*{\large Data availability}
\label{sec:data_n}
All data that support the findings of this study are included in the article and the Supplementary Information file. These data are available from the corresponding author upon request.
\justify

\section*{\large Code availability}
\label{sec:code_n}
All the codes that support the findings of this study are available from the corresponding authors upon request.

\bibliography{ref.bib}

\bibliographystyle{Nature}
\section*{Acknowledgements}

This work is partially supported by NSF 2246149, 2344819. Experimental validation is supported by the U.S. Department of Energy, Office of Science,
Office of Basic Energy Sciences Energy Frontier Research Centers program under Award
Number DESC0021118. 
\label{sec:acknowledgements_n}
\section*{Authors Contributions}

S.G., K.N., T.K. proposed and supervised the project. S.A.O. and M.A.I.R. conducted the cell design, SPICE verification, and system benchmarking. J.D., F.M., and T.K. performed experimental validation. All authors contributed to write up of the manuscript.

\label{sec:authors_contributions_n}
\section*{Competing Interests}

The authors declare that they have no competing interests.
\label{sec:competing_n}



\clearpage 

\begingroup
  \let\clearpage\relax
  \let\cleardoublepage\relax
  \let\newpage\relax

  \setcounter{figure}{0}
  \renewcommand{\thefigure}{S\arabic{figure}}

  \renewcommand{\thefigure}{S\arabic{figure}}
\renewcommand{\thetable}{S\arabic{table}}
\justify

\onecolumn
\centering
\textbf{\Large Supplementary Information}
\setcounter{table}{0}
\setcounter{page}{1}
\begin{flushleft}

\subsection*{Device information}
The AND-type FeFET arrays were fabricated in GlobalFoundries’ 28 nm HKMG technology. This process allows seamless integration of CMOS transistors and FeFETs within a single chip. The FeFET structure features a 8 nm Si-doped HfO\textsubscript{2} ferroelectric layer and a 1 nm SiO\textsubscript{2} interface layer. All the FeFETs in this array have a $W/L = 0.5\mu m/0.5\mu m$.

\subsection*{Electrical characterization}
The electrical characterization was performed on a AND-type FeFET array, which consisted of 8 wordlines (WLs) connecting the gates row-wise and 6 bitlines/sourcelines (BLs/SLs) connecting the drain and source contacts column-wise. All FeFETs are preconditioned for 50 program-erase cycles with write pulses of 4.5 V 500 ns and -5 V 500 ns. The pulses are provided to the FeFET array via pulse generators and pulse shaping boards as given in Fig. \ref{fig:Experimental Setup}. The FeFET read operation is performed by applying a voltage ramp from -0.2 V to 1.8 V in increments of 100 mV while measuring the currents at the drain electrode biased at 100 mV with the bulk and source grounded. All the unselected WLs are biased at -0.5 V to suppress leakage currents.

\begin{figurehere}
	\centering
	\includegraphics [width=0.6\textwidth,keepaspectratio]{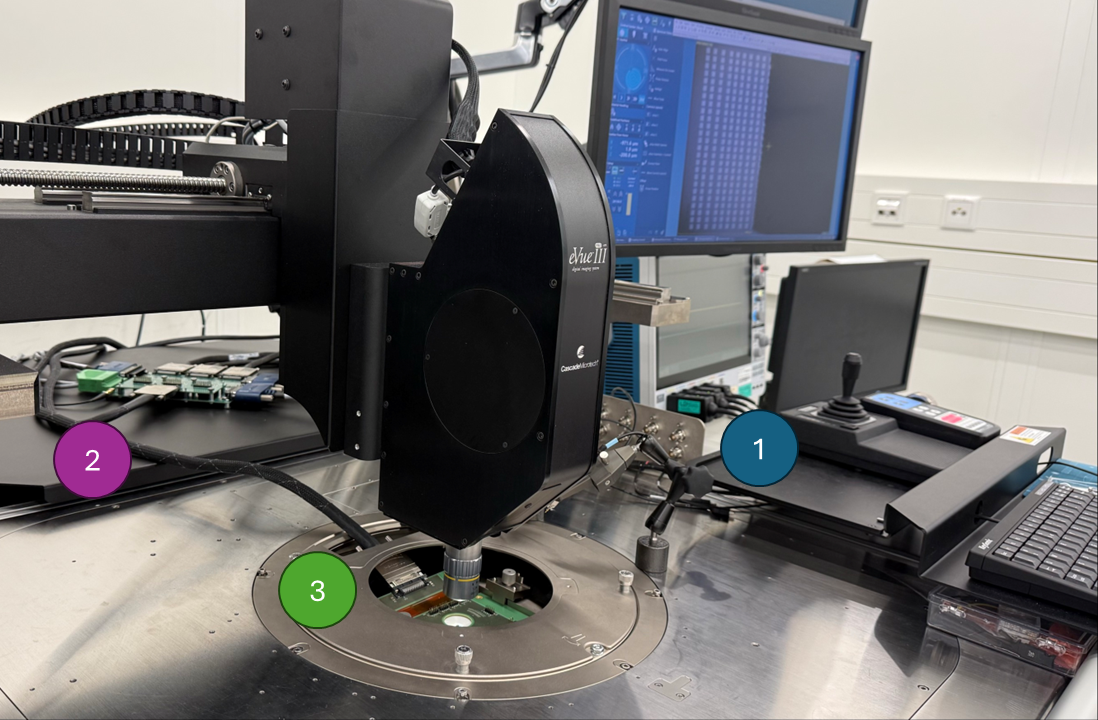}
	\caption{Experimental wafer-level probing setup for the FeFET array testing with 1) the recording system incl. pattern generator, SMU; 2) the adapter board for pulse shaping; 3) needle card connecting the FeFET wafer.}
	\label{fig:Experimental Setup}
\end{figurehere}

\end{flushleft}
\justify

\subsection*{Encrypted Array Architecture and Decryption }

In the 1‑bit FeFET encrypted memory array shown in Fig.~\ref{fig:array_read}, decryption of one row of ciphertext (CT) can be completed in a single cycle. The memory array shares the wordline (WL) along each row, while both the bitline (BL) and sourceline (SL) are shared along each column. Fine-grained keys are applied as bias voltages on the vertical BL/SL lines, whereas the read voltage $V_R$ is applied to the shared WL of the selected row. Consequently, both key values (0 and 1) can be applied in parallel across different columns in this array architecture. In contrast, prior work~\cite{embedding_security_into_ferroelectric_fet_array_via_in_situ_memory_operation} applies the key through shared WL lines and therefore requires two decryption cycles per row. Fig.~\ref{fig:array_read}(a–d) illustrates plaintext (PT) recovery in a $2\times2$ array of ciphertexts with a column-wise key set of $(0,1)$. Row~1 stores the ciphertext $(0,0)$, which corresponds to the device configuration (LVT, LVT). As shown in Fig.~\ref{fig:array_read}(a), to apply key $=0$ to the first ciphertext column, the bias $(BL_1/SL_1) = (\mathrm{GND}/V_{\mathrm{DD}})$ is applied. Similarly, to apply key $=1$ to the second ciphertext column, the bias $(BL_2/SL_2) = (V_{\mathrm{DD}}/\mathrm{GND})$ is used. When $V_R$ is asserted on $WL_1$, both FeFETs in Row~1 are LVT devices; thus, both SL lines are driven to the voltage level supplied at their respective BLs, as shown in Fig.~\ref{fig:array_read}(b). Specifically, $SL_1$ is pulled to GND and $SL_2$ is pulled to $V_{\mathrm{DD}}$, yielding plaintext $(0,1)$, consistent with the XOR relation $PT = Key \oplus CT$, i.e., $Key=(0,1)$ and $CT=(0,0)$.
Fig.~\ref{fig:array_read}(c) shows the application of the same column-wise key pattern $(0,1)$ to the stored ciphertext $(0,1)$ in Row~2, which corresponds to the device configuration (LVT, HVT). In this case, when $V_R$ is applied to WL2, $SL_1$ is driven to GND because the corresponding FeFET is LVT. However, the FeFET in the second column is HVT, so $SL_2$ remains at GND. As a result, the final SL voltages for the two columns are $(\mathrm{GND}, \mathrm{GND})$, which translates to plaintext $(0,0)$. This outcome again matches the XOR logic $PT = Key \oplus CT$, with $Key=(0,1)$ and $CT=(0,1)$.

\begin{figurehere}
\centering
\vspace{1 mm}
\includegraphics[width=1\textwidth]{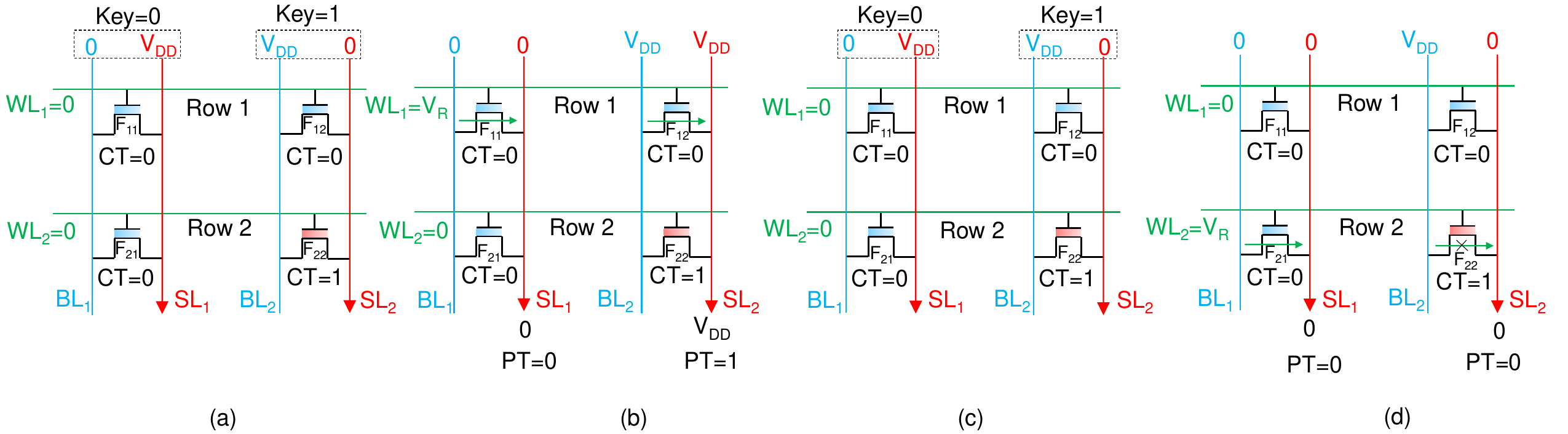}
\caption{Encrypted memory array architecture and single-row decryption. 
(a) Application of $\mathrm{Key} = (0,1)$ to $\mathrm{CT} = (0,0)$ in Row~1 for decryption. (b) Sensed SL voltages corresponding to $\mathrm{PT} = (0,1)$. (c) Application of $\mathrm{Key} = (0,1)$ to $\mathrm{CT} = (0,1)$ in Row~2 for decryption. (d) Sensed SL voltages corresponding to $\mathrm{PT} = (0,0)$.}


\label{fig:array_read}
\end{figurehere}

\textbf{FeFET Programming}

The FeFET array writing is illustrated in Fig.~\ref{fig:writing implementation}. The target array state is CT = (0,1) in the first row, corresponding to (LVT, HVT). FeFETs  in a row are written using a two-step procedure following the $V_W/3$ scheme~\cite{Write_Disturb_in_Ferroelectric_FETs_and_Its_Implication_for_1T-FeFET_AND_Memory_Arrays}. In the first cycle, all cells are initialized to HVT by applying $-V_W$ to each WL. In the second cycle, the target FeFET $F_{11}$ is programmed to LVT by applying $+V_W$ on $WL_1$. To inhibit a change in polarization of the other cell $F_{12}$, a voltage of $2V_W/3$ is applied to both $BL_2$ and $SL_2$. The FeFET $F_{12}$ retains its polarization state because the gate-to-source voltage $V_{gs}$ is only $V_W/3$, which is insufficient to switch the polarization. Also, to inhibit the polarization switching in $F_{22}$, a voltage of $V_W/3$ is applied to $WL_2$, at the second step in the unselected row.

{
\makeatletter
\renewcommand{\thefigure}{S\arabic{figure}}
\renewcommand{\fnum@figure}{Fig.\ \thefigure}
\makeatother
\begin{figurehere}
\centering
\includegraphics[width=1\textwidth,keepaspectratio]{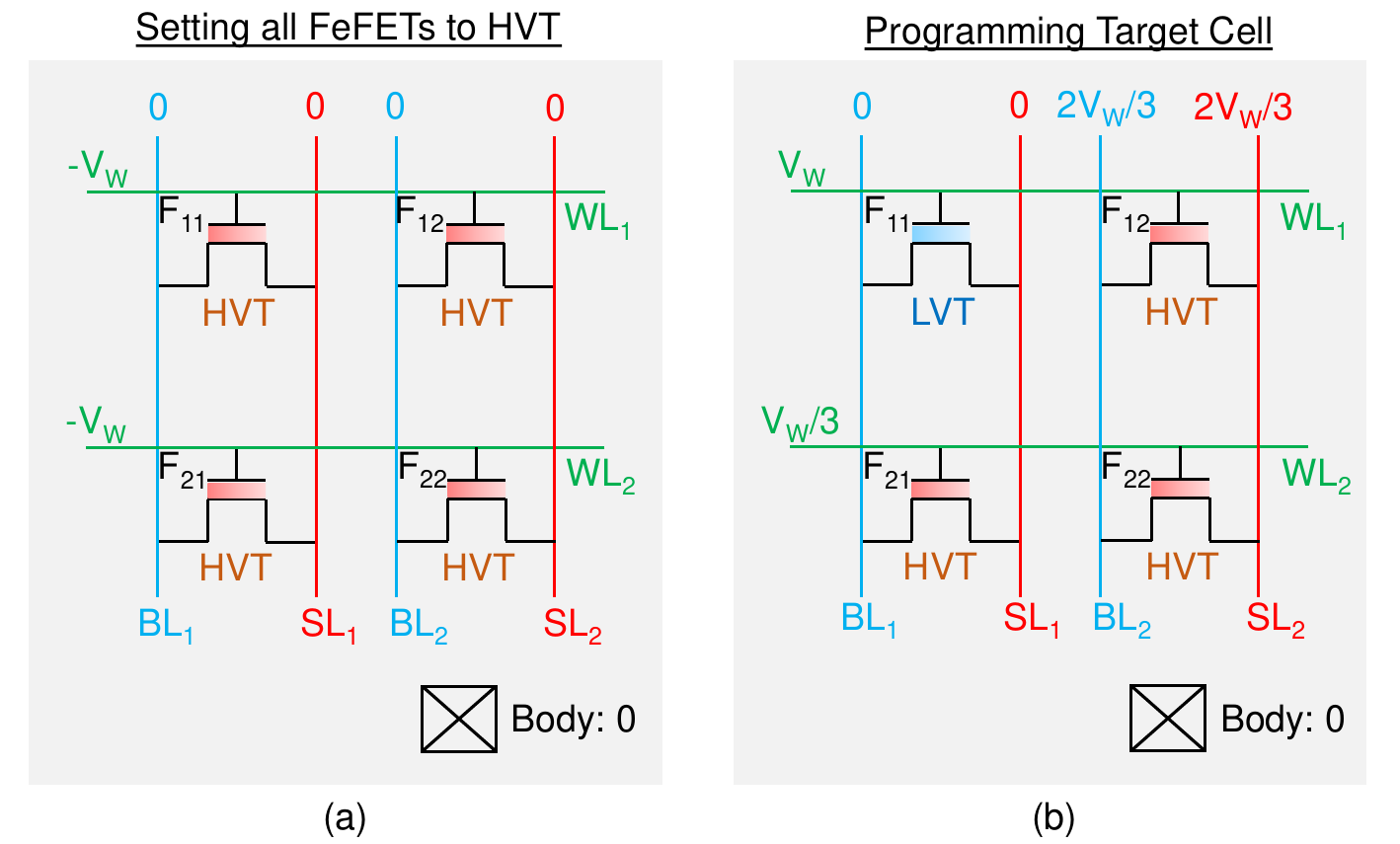}
    
    \caption{Proposed two-step programming scheme. (a) First step: all cells are reset to the high-threshold-voltage (HVT) state; (b) second step: the target cells in the selected row are programmed to the low-threshold-voltage (LVT) state while inhibition biasing is applied to all unselected cells.\cite{embedding_security_into_ferroelectric_fet_array_via_in_situ_memory_operation} \cite{Write_Disturb_in_Ferroelectric_FETs_and_Its_Implication_for_1T-FeFET_AND_Memory_Arrays}.}
\label{fig:writing implementation}
\end{figurehere}
}
\subsection*{Simulation Verification}
 
\begin{figurehere}
\centering
\vspace{1 mm}
\includegraphics[width=0.80\textwidth]{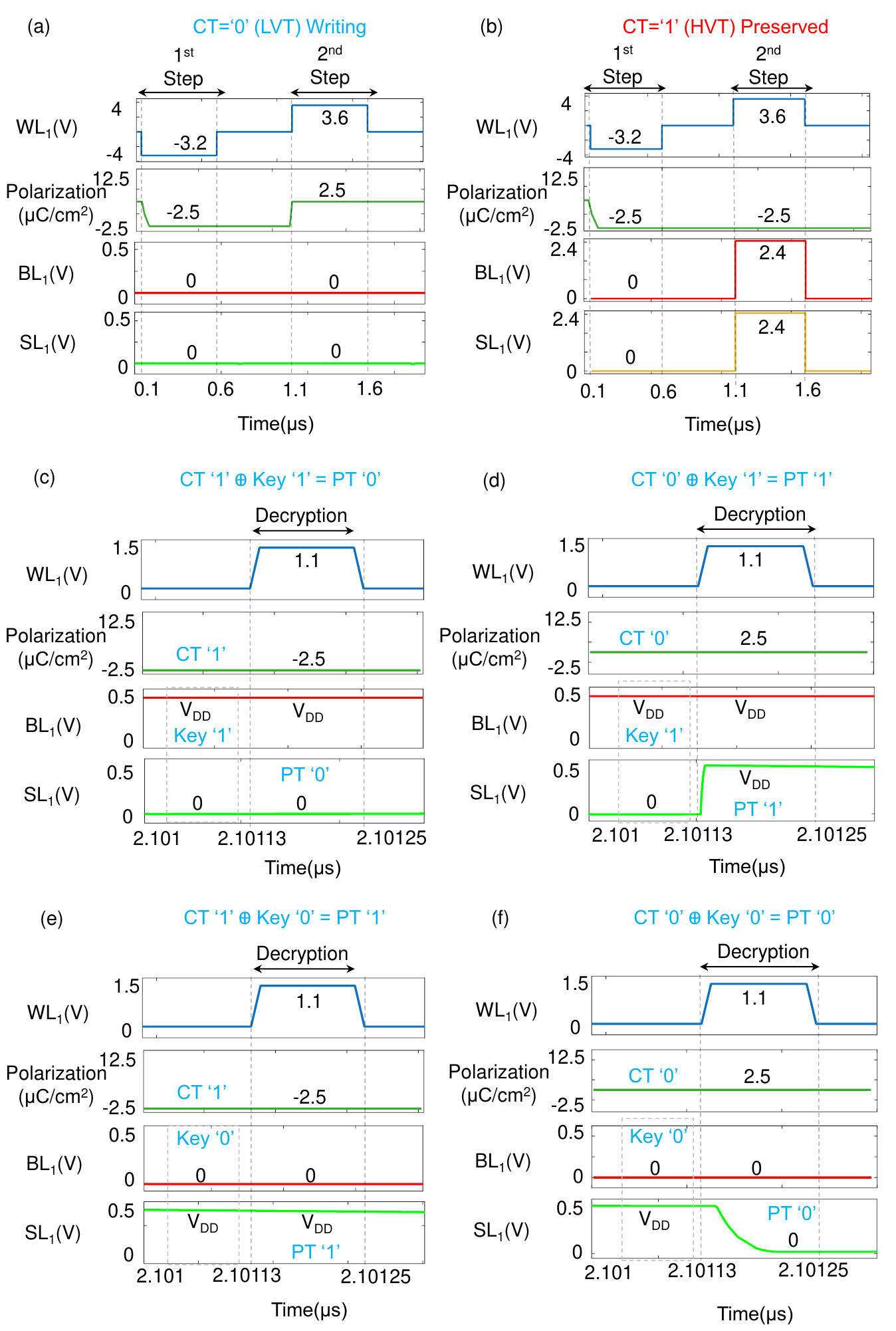}
\caption{Transient response of the FeFET programming and XOR-based decryption. (a) Programming of the FeFET cell F\textsubscript{11} to CT '1' (HVT)  first and then to CT '0' (LVT) second; (b) FeFET cell F\textsubscript{11} remaining at CT '1' (HVT) under inhibition biasing. Decryption process: (c) CT '1' $\oplus$ Key '1' = PT '0'; (d) CT '0' $\oplus$ Key '1' = PT '1'; (e) CT '1' $\oplus$ Key '0' = PT '1'; (f) CT '0' $\oplus$ Key '0' = PT '0'.}
\label{fig:Cadence Waveform}
\end{figurehere}

\justify
Simulation and analysis are performed in the Cadence Virtuoso  ADE-XL  Spectre environment using the open-source NCSU 45\,nm Basekit~\cite{NCSU_Basekit} together with a multi-domain FeFET compact model~\cite{A_Circuit_Compatible_Accurate_Compact_Model_for_Ferroelectric-FETs}. The functionality of the proposed design is verified and discussed in detail in this section based on the array structure demonstrated in Fig.~\ref{fig:array_read}. Fig.~\ref{fig:Cadence Waveform} illustrates the transient simulation results for encrypted FeFET cell programming (refer Fig.~\ref{fig:writing implementation}).
The programming of ciphertext logic `0` and `1` is shown in Fig.~\ref{fig:Cadence Waveform}(a) and Fig.~\ref{fig:Cadence Waveform}(b), respectively. A negative gate bias of 3.2\,V ($-V_W$) is applied to reset the FeFET polarization to $-2.5 \,\mu\text{C}/\text{cm}^2$ (HVT), whereas a positive gate bias of 3.6\,V ($+V_W$) sets the polarization to  $2.5 \,\mu\text{C}/\text{cm}^2$ (LVT), with both drain and source terminals grounded. A $V_W/3$ write scheme is adopted to prevent unselected cells from switching their polarization states~\cite{Write_Disturb_in_Ferroelectric_FETs_and_Its_Implication_for_1T-FeFET_AND_Memory_Arrays}. During the second step of the programming operation, unselected BL and SL lines are biased at 2.4\,V ($+2V_W/3$), while all unselected WL lines are biased at 1.2\,V ($+V_W/3$). This biasing condition preserves the polarization state of unselected cells even when the programming pulse is applied to the selected row.

Fig. \ref{fig:Cadence Waveform}(c-f) covers the decryption process of all 4 possible combinations of 1-bit ciphertext and key. After the SL is precharged and the appropriate key-dependent bias voltage is applied to the BL, the plaintext data are observed at the SL when a read voltage V\textsubscript{R} is applied to the WL, with V\textsubscript{LVT} $<$ V\textsubscript{R} $<$ V\textsubscript{HVT}. A read voltage of V\textsubscript{R} = 1.1 V is chosen, which enables approximately 98\% drain-to-source voltage transfer when the FeFET is in the LVT state. Figure \ref{fig:Cadence Waveform}(c) illustrates decryption of ciphertext CT = 1 (HVT) using key = 1 with BL/SL = (V\textsubscript{DD}/ GND) and V\textsubscript{DD} = 0.5 V. When V\textsubscript{R} is applied to the WL, the SL remains at ground, corresponding to plaintext PT = 0. Figure \ref{fig:Cadence Waveform}(d) shows decryption of ciphertext CT = 0 (LVT) using the same key = 1 and BL/SL = (V\textsubscript{DD}/ GND). In this case, when V\textsubscript{R} is applied to the WL, the SL rises close to V\textsubscript{DD} through the conducting LVT FeFET, indicating plaintext PT = 1. Figure \ref{fig:Cadence Waveform}(e) demonstrates decryption of ciphertext CT = 1 (HVT) using key = 0 with BL/SL = (GND/ V\textsubscript{DD}). Upon application of V\textsubscript{R} to the WL, the HVT FeFET remains off, so the SL stays at V\textsubscript{DD}  corresponding to plaintext PT = 1. As shown in Figure \ref{fig:Cadence Waveform}(f), decryption of ciphertext CT = 0 (LVT) with key = 0 and BL/SL = (GND/ V\textsubscript{DD}) causes the SL to be discharged to ground when V\textsubscript{R} is applied to the WL, indicating plaintext PT = 0.

\end{flushleft}


\begin{flushleft} 
\textbf{\large MLC XOR Decryption Simulation}
\end{flushleft}
Figure~\ref{fig:mlc_xor} illustrates four example combinations of 2-bit ciphertext and 2-bit key based on the array structure in Fig.~\ref{fig:array_read}. Four polarization states of 2.5, 0.0, $-2.5$, and $-5.0~\mu\text{C}/\text{cm}^2$ are selected for $\mathrm{CT} = 00, 01, 10,$ and $11$, respectively. 
In the first read step, BL and SL are biased according to $\mathrm{Key}_{\mathrm{MSB}}$, and the WL is driven with 1.8~V ($V_{\text{R2}}$). The resulting SL voltage is sensed to determine $\mathrm{PT}_{\mathrm{MSB}}$. In the second read step, $\mathrm{Key}_{\mathrm{LSB}}$ is applied to the BL and SL, and a read pulse of 1.1~V ($V_{\text{R1}}$) is applied to the WL. The BL and SL are then rebiased with $\mathrm{Key}_{\mathrm{LSB}}$, and a third read pulse of 2.5~V ($V_{\text{R3}}$) is applied to the WL in the third step. Depending on $\mathrm{CT}_{\mathrm{MSB}}$, the SL voltage in either the second or third cycle determines $\mathrm{PT}_{\mathrm{LSB}}$.

Figure~\ref{fig:mlc_xor}(a) shows decryption when $(\mathrm{CT}, \mathrm{Key}) = (11,11)$. The level $\mathrm{CT} = 11$ corresponds to the highest threshold voltage and therefore does not conduct for any of the three read pulses. For $\mathrm{Key} = 11$, the BL/SL pair is biased as $(V_{\text{DD}}, \mathrm{GND})$ in all three cycles, so the recovered plaintext is $\mathrm{PT} = 00$.
Figure~\ref{fig:mlc_xor}(b) shows decryption when $(\mathrm{CT}, \mathrm{Key}) = (00,00)$. The level $\mathrm{CT} = 00$ has the lowest threshold voltage and conducts for all three read pulses. With $\mathrm{Key} = 00$, the BL/SL pair is biased as $(\mathrm{GND}, V_{\text{DD}})$ in each cycle, causing the SL to be pulled to GND every time, and the plaintext is decoded as $\mathrm{PT} = 00$.
Figure~\ref{fig:mlc_xor}(c) illustrates decryption for $(\mathrm{CT}, \mathrm{Key}) = (10,01)$. The MSB of $\mathrm{Key} = 01$ is encoded as $\mathrm{BL}/\mathrm{SL} = (\mathrm{GND}, V_{\text{DD}})$ in the first step. The threshold voltage associated with $\mathrm{CT} = 10$ is such that the FeFET turns on only when the highest read voltage $V_{\text{R3}}$ is applied; consequently, it does not conduct in the first step when the WL is driven with 1.8~V, and the SL remains close to $V_{\text{DD}}$. The LSB of $\mathrm{Key} = 01$ is encoded as $\mathrm{BL}/\mathrm{SL} = (V_{\text{DD}}, \mathrm{GND})$ in the second and third cycles. For $\mathrm{CT} = 10$, the second cycle is irrelevant. In the third cycle, when $V_{\text{R3}} = 2.5$~V is applied to the WL, the FeFET conducts and the SL is driven close to $V_{\text{DD}}$. Overall, this sequence decodes the plaintext as $\mathrm{PT} = 11$.
Figure~\ref{fig:mlc_xor}(d) considers $\mathrm{CT} = 01$, which corresponds to the second-lowest threshold voltage, and $\mathrm{Key} = 11$. In all cycles, the BL/SL pair is biased as $(V_{\text{DD}}, \mathrm{GND})$. In the first cycle, the applied read voltage causes the FeFET to conduct and the SL to rise close to $V_{\text{DD}}$. In the second cycle, when the WL is driven with $V_{\text{R1}} = 1.1$~V, the threshold voltage is higher than $V_{\text{R1}}$, so the FeFET is off and the SL remains at GND. The combination of SL levels across the read cycles corresponds to a decoded plaintext of $\mathrm{PT} = 10$.

{
\makeatletter
\renewcommand{\thefigure}{S\arabic{figure}}
\renewcommand{\fnum@figure}{Fig.\ \thefigure}
\makeatother

\begin{figurehere}
\centering
\includegraphics[width=1\textwidth]{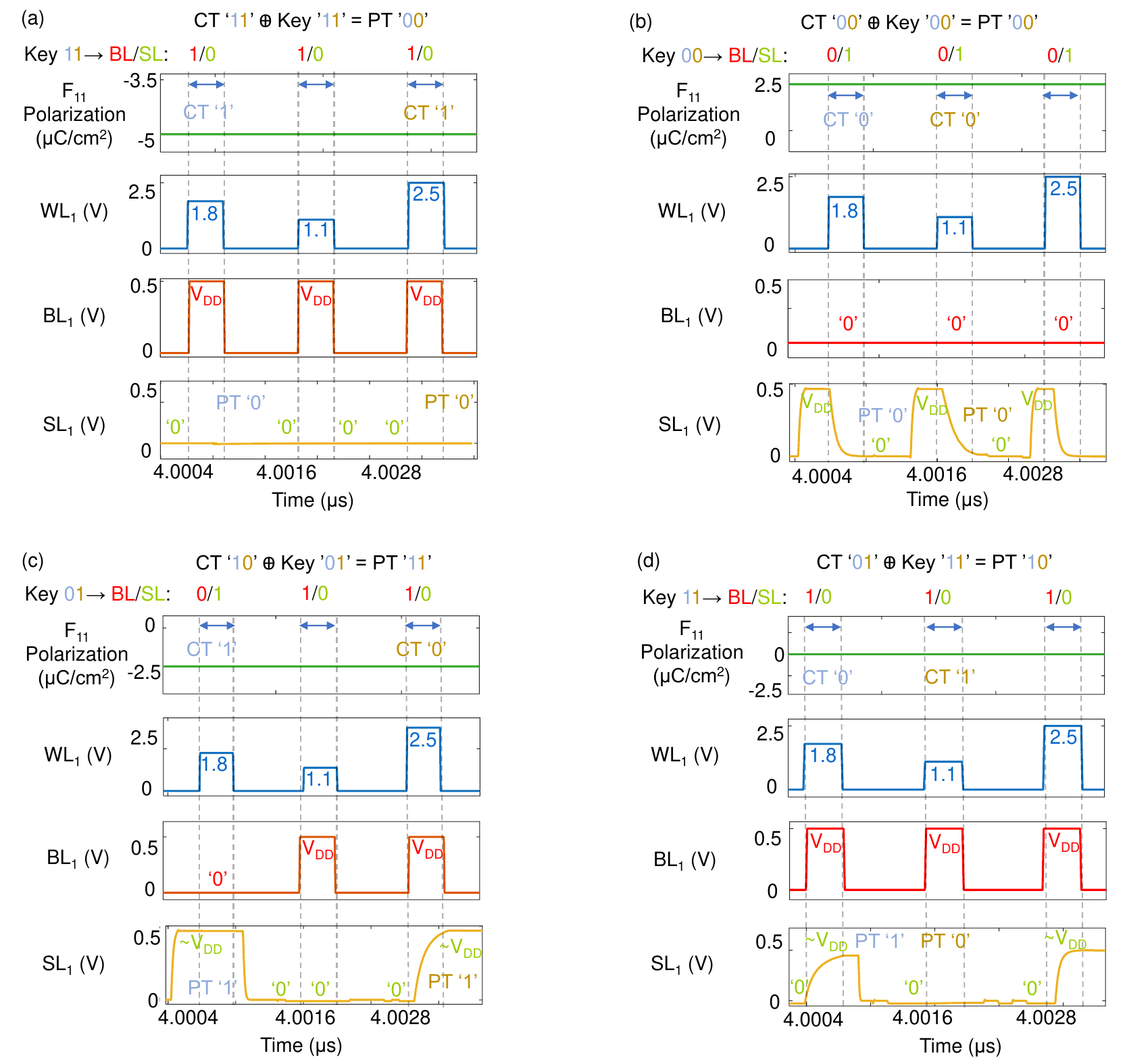}
\caption{Transient response of proposed FeFET-based MLC XOR decryption, (a) CT '11' $\oplus$ Key '11' = PT '00', (b) CT '00' $\oplus$ Key '00' = PT '00', (c) CT '10' $\oplus$ Key '01' = PT '11',(d) CT '01' $\oplus$ Key '11' = PT '10' }
\label{fig:mlc_xor}
\end{figurehere}
}

\begin{flushleft} 
\textbf{\large Sensitivity anslysis: Monte Carlo Analysis}
\end{flushleft}

To evaluate the read sensitivity of the proposed decryption scheme, we performed a 1000-point Monte Carlo (MC) sweep in which the transistor threshold voltage was perturbed with a standard deviation of $\sigma=\SI{40}{\milli\volt}$~\cite{First_demonstration_of_in-memory_computing_crossbar_using_multi-level_Cell_FeFET}. The output was measured at the source line (SL), while the access gate was driven with a \SI{1.1}{\volt}, \SI{100}{\pico\second} read pulse, and the SL voltage was sampled at $t=\SI{100}{\pico\second}$. All simulations were carried out in Cadence Virtuoso, based on the model \cite{NCSU_Basekit} and the resulting distributions are summarized in Fig.~\ref{fig:monte_carlo}. Figure~\ref{fig:monte_carlo}(a) shows the threshold-voltage distributions corresponding to the low-$V_{\mathrm{th}}$ (LVT: 0.4V) and high-$V_{\mathrm{th}}$ (HVT:1.75V) device states. The resulting SL-voltage ($V_{\mathrm{SL}}$) distributions across the 1000 MC samples are presented in Fig.~\ref{fig:monte_carlo}(b), revealing a worst-case sense margin of \SI{0.245}{\volt} between PT~=~`0' and PT~=~`1'. Four combinations of CT and Key are illustrated. When CT~=~`1' (HVT), both Key states---Key~=~`1' with $(V_{\mathrm{BL}},V_{\mathrm{SL}})=(\SI{0.5}{\volt},\SI{0}{\volt})$ and Key~=~`0' with $(V_{\mathrm{BL}},V_{\mathrm{SL}})=(\SI{0}{\volt},\SI{0.5}{\volt})$---yield narrow $V_{\mathrm{SL}}$ distributions, as the transistor is effectively OFF. In contrast, when CT~=~`0' (LVT), the device is ON for both key conditions, leading to broader $V_{\mathrm{SL}}$ distributions.

\begin{flushleft}

{
\makeatletter
\renewcommand{\thefigure}{S\arabic{figure}}
\renewcommand{\fnum@figure}{Fig.\ \thefigure}
\makeatother
\begin{figurehere}
\centering
\includegraphics[width=1\textwidth]{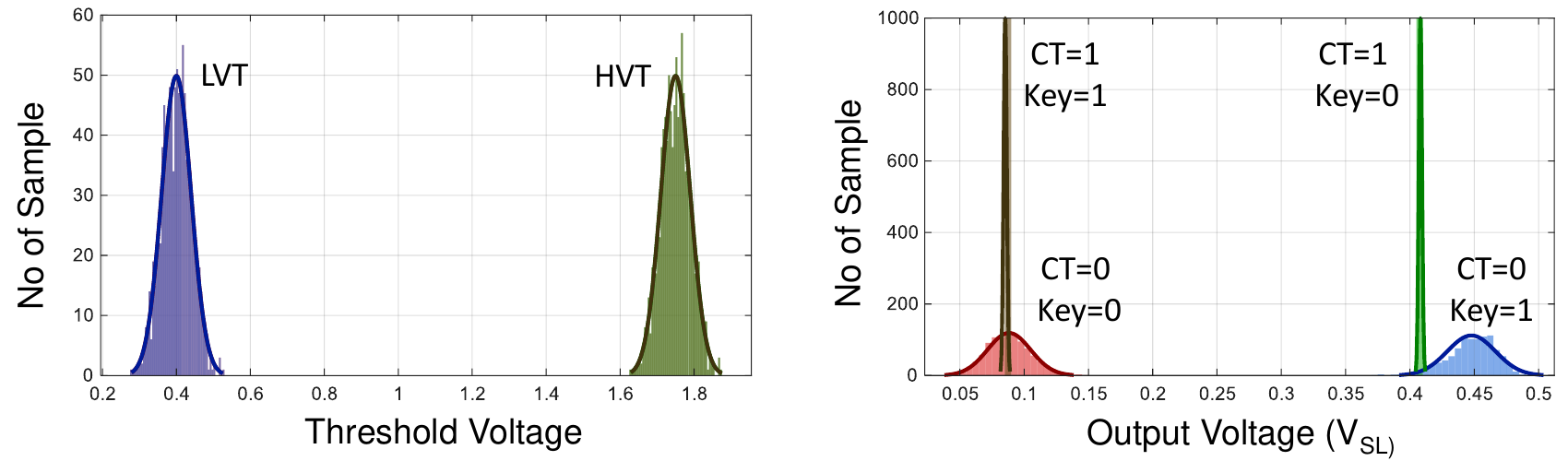}
\caption{(a) V\textsubscript{th}-distribution for 1000 Monte~Carlo samples,(b) V\textsubscript{SL} (PT)voltage distribution for four different combinations of CT \& Key.}
\label{fig:monte_carlo}
\end{figurehere}
}
\textbf{\large Layout}
\end{flushleft}

Fig.\ref{fig:layout} shows the layout of the $2 \times 2$ array of the proposed 1T-encryption. For design  GPDK045  rules \cite{cadence_gpdk45} has been used. The area footprint of the $2 \times 2$ array is 0.7912 $\mu m^2$ for 45nm technology node.  

{
\makeatletter
\renewcommand{\thefigure}{S\arabic{figure}}
\renewcommand{\fnum@figure}{Fig.\ \thefigure}
\makeatother
\begin{figurehere}
\centering
\includegraphics[width=1\textwidth]{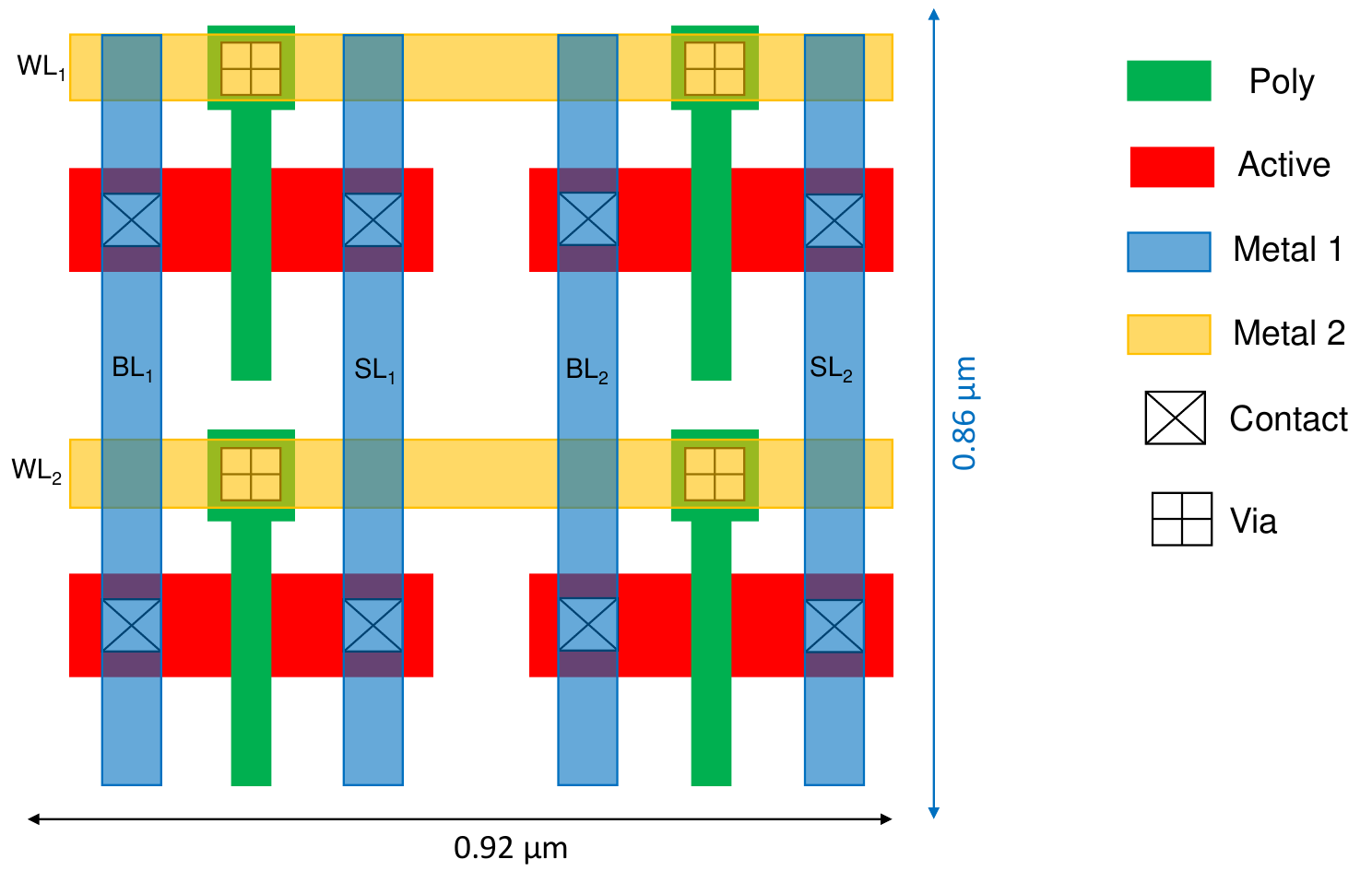}
\caption{2X2 array layout of proposed encryption/decryption}
\label{fig:layout}
\end{figurehere}
}
\endgroup


\end{document}